\documentclass[preprint,groupedaddress,tightenlines,nofootinbib,floatfix,dvitex]{revtex4-1}

\usepackage{graphicx}
\usepackage{dcolumn}
\usepackage{amssymb}
\usepackage{amsmath}
\usepackage{epstopdf}
\usepackage{xcolor}         

\makeatletter
\newcommand{\figcaption}[1]{\def\@captype{figure}\caption{#1}}
\newcommand{\tblcaption}[1]{\def\@captype{table}\caption{#1}}
\makeatother

\def\simge{\mathrel{%
       \rlap{\raise 0.511ex \hbox{$>$}}{\lower 0.511ex \hbox{$\sim$}}}}
\def\simle{\mathrel{%
       \rlap{\raise 0.511ex \hbox{$<$}}{\lower 0.511ex \hbox{$\sim$}}}}

\begin{document}

\title{Phase structure of heavy dense lattice QCD and the three-state Potts model}

\author{Shinji Ejiri$^{1}$ and Masanari Koiida$^{2}$}

\affiliation{
$^{1}$Department of Physics, Niigata University, Niigata 950-2181, Japan\\
$^{2}$Graduate School of Science and Technology, Niigata University, Niigata 950-2181, Japan}

\date{March 13, 2026}

\begin{abstract}
The nature of the finite temperature phase transition of QCD depends on the particle density and the mass of the dynamical quarks. 
We discuss the properties of the phase transition at high density, considering an effective theory describing the high-density heavy-quark limit of QCD.
This effective theory is a simple model in which the Polyakov loop is a dynamical variable, and the quark Boltzmann factor is controlled by only one parameter, $C(\mu,m_q)$, which is a function of the quark mass $m_q$ and the chemical potential $\mu$. 
The Polyakov loop is an order parameter of $Z_3$ symmetry, and the fundamental properties of the phase transition are thought to be determined by the $Z_3$ symmetry broken by the phase transition.
By replacing the Polyakov loop with $Z_3$ spin, we find that the effective model becomes a three-dimensional three-state Potts model ($Z_3$ spin model) with a complex external field term.
We investigate the phase structure of the Potts model and discuss QCD in the heavy-quark region.
As the density varies from $\mu=0$ to $\mu=\infty$, we find that the phase transition is first order in the low-density region, changes to a crossover at the critical point, and then becomes first-order again. 
This strongly suggests the existence of a first-order phase transition in the high density heavy-quark region of QCD.
\end{abstract}

\maketitle

\section{Introduction}
\label{sec:intro}

The finite temperature phase transition of QCD that occurred immediately after the birth of the Universe is thought to be a crossover without thermodynamic singularities.
However, the nature of the phase transition depends on the particle density and the mass of the dynamical quark.
It is well known that in the case of infinite quark mass, i.e., quenched QCD, the finite temperature phase transition is first order, and as the mass decreases the phase transition changes to a crossover at a critical mass, and the critical point belongs to the three-dimensional Ising universality class \cite{Saito:2011fs,Saito:2013vja,Ejiri:2019csa,Kiyohara:2021smr,Ashikawa:2024njc,Cuteri:2020yke}.
There has been a great deal of effort in studying the phase transition in the light-quark region, i.e., near the chiral limit \cite{Bazavov:2017xul,Jin:2017jjp,HotQCD:2018pds,Kuramashi:2020meg,Borsanyi:2020fev,Cuteri:2021ikv,Dini:2021hug,Borsanyi:2021hbk,Gavai:2024mcj,Zhang:2025vns}.
Recently, there has been a lot of discussion about whether the finite temperature phase transition in the three-flavor chiral limit is first order in the continuum limit, but the standard understanding is that in the massless limit, the phase transition in two-flavor QCD is second order, and in the case of three or more flavors, the phase transition is first order \cite{Pisarski:1983ms}.
At least for finite lattice spacing, there is a first-order transition region near the three-flavor quark massless limit.

Numerical simulations of lattice QCD have shown that in the heavy-quark region, the critical mass at which a first-order phase transition turns into a crossover increases with increasing density \cite{Saito:2013vja,Ejiri:2023tdp}.
When investigating high-density regions in lattice QCD simulations, a problem known as the sign problem arises, making it difficult to study. 
However, in regions where quarks are sufficiently heavy, the sign problem is relatively less serious, making research possible.
On the other hand, the critical mass at which the first-order phase transition turns into a crossover in the light-quark region is also expected to increase with increasing density, and various studies have been conducted to confirm this \cite{Karsch:2003va,deForcrand:2003vyj,Ejiri:2012rr,Jin:2013wta}.
However, the sign problem in the light-quark region is serious, and research in this region has not progressed very far.
If the critical mass increases, it is expected that the QCD phase transition will change from a crossover to a first-order phase transition as density increases, even for real quark masses, which is very interesting.

Furthermore, if the first-order phase transition region for light quarks continues to expand with increasing density, the critical point should reach the studiable heavy-quark region.
In this study, we discuss whether there is a first-order phase transition region in the heavy-quark high-density region that is separate from the first-order phase transition in the heavy-quark low-density region.

The fundamental properties of a phase transition are thought to be determined solely by the symmetry broken during the phase transition and the dimension of the space.
In the early days of finite temperature lattice QCD, the three-state Potts model \cite{Wu:1982ra}, which has the same broken symmetry, was used to discuss how the nature of the phase transition changes when the effects of dynamical quarks are added from quenched QCD \cite{Banks:1983me,DeGrand:1983fk}.
When the quark mass is sufficiently large, the hopping parameter expansion leads to the effective action of the quark, whose leading term is proportional to the Polyakov loop.
The Polyakov loop is the order parameter of the $Z_3$ center symmetry that is broken at the finite temperature phase transition of QCD.
In comparison with the three-state Potts model, which also experiences a phase transition due to $Z_3$ symmetry breaking, it has been argued that the first-order phase transition in quenched QCD turns into a crossover due to the effects of dynamical quarks.
The Polyakov loop corresponds to the $Z_3$ spin, and the quark determinant corresponds to the external magnetic field term in the spin model.
In both models, the critical point where the first-order phase transition turns into a crossover is known to belong to the Ising universality class.

In this study, we apply this argument, which explores the nature of the phase transition in comparison with the spin model, to high-density lattice QCD.
There is an effective theory that describes the heavy-quark high-density limit of QCD \cite{Blum:1995cb,Aarts:2008rr,Aarts:2017vrv}.
The effective theory is a simple model in which the Polyakov loop is the fundamental dynamical variable and the Boltzmann factor of a quark is controlled by only one parameter $C(\mu, m_q)$, which is a function of the quark mass $m_q$ and the chemical potential $\mu$.
We show that by replacing the Polyakov loop with a $Z_3$ spin, the effective model becomes a three-dimensional three-state Potts model ($Z_3$ spin model) with a complex external magnetic field term.
We investigate the phase structure of the Potts model corresponding to QCD in the heavy-quark region and how the nature of the phase transition changes when the chemical potential is increased.
In particular, we discuss whether there is a first-order phase transition in the heavy-quark high-density region of QCD.

Although not discussed in this paper, the three-dimensional Potts model can be analyzed using the tensor renormalization group \cite{Xie:2012mjn,Morita:2018tpw}.
The tensor renormalization group method is independent of the sign problem and is expected to be effective for calculations of systems with sign problems.
However, the difficulty of applying it to higher-dimensional models is a weakness of the method.
Therefore, the use of the tensor renormalization group is also one of our motivations for constructing a three-dimensional effective theory.

The paper is written as follows: in Sec.~\ref{sec:hdqcd}, we describe the heavy-quark high-density effective theory of lattice QCD and the corresponding spin model.
We also explain the high-density/low-density duality in the effective theory of heavy dense QCD, which is important in this research.
In Sec.~\ref{sec:critical}, we determine the critical point where the first-order phase transition ends in the three-state Potts model, which corresponds to heavy dense QCD.
The crossover region is discussed in Sec.~\ref{sec:crossover}.
In Sec.~\ref{sec:complexmu}, we argue for the case of extending the chemical potential to complex numbers.
Section \ref{sec:summary} is devoted to conclusions and future perspectives.

\section{Heavy dense QCD and three-state Potts model}
\label{sec:hdqcd}

\subsection{Hopping parameter expansion and the effective parameter}

In the heavy-quark region, we evaluate the quark determinant by the hopping parameter expansion around $\kappa=0$,
\begin{eqnarray}
\ln \det M(\kappa) = 
\ln \det M(0) + N_{\rm site} \sum_{n=1}^{\infty} D_{n} \kappa^{n} .
\label{eq:tayexp} 
\end{eqnarray}
These $D_n$ can be decomposed into the number of windings $m$ in the time direction,
\begin{eqnarray}
D_n &=& \hat{W}(n) + \sum_{m=1}^{\infty} \hat{L}_m^+ (N_t, n) e^{m \mu/T} + \sum_{m=1}^{\infty} \hat{L}_m^- (N_t, n) e^{-m \mu/T}
\nonumber \\
&=&  \hat{W}(n) + \sum_{m=1}^{\infty} 2 {\rm Re} \hat{L}_m^+ (N_t, n) \cosh \left( \frac{m \mu}{T} \right) 
+ i \sum_{m=1}^{\infty} 2 {\rm Im} \hat{L}_m^+ (N_t, n) \sinh \left( \frac{m \mu}{T} \right) .
\label{eq:loopex}
\end{eqnarray}
In this expansion, the effect of the chemical potential appears as above.
Nonzero contributions appear when the products of the hopping terms form closed loops in the spacetime. 
Therefore, the nonzero contributions to $D_n$ are given by $n$-step Wilson loops and Polyakov loops. 
The latter are closed by the boundary condition, where we impose the antiperiodic boundary condition in the time direction for fermions.
$\hat{W} (n)$ is the sum of Wilson loop-type terms with winding number zero, and $\hat{L}_m^+ (N_t, n)$ is the sum of Polyakov loop-type terms with winding number $m$ in the positive direction. 
$\hat{L}_m^- (N_t, n)$ is that with winding number $m$ in the negative direction, 
and $\hat{L}_m^- (N_t, n)=[\hat{L}_m^+ (N_t, n)]^*$.
$\hat{W}(n)$ is nonzero only if $n \geq 4$ and $n$ is even.
When $N_t$ is an even number, $\hat{L}_m^+ (N_t, n)$ is nonzero only if $n$ is an even number, and the lowest-order term is $\hat{L}_m^+ (N_t, mN_t)$, i.e.,
$\hat{L}_m^+ (N_t, n)$ can be nonzero when $n \geq m N_t$.

These expansion coefficients have been calculated on configurations generated near the phase transition point in Ref.~\cite{Wakabayashi:2021eye} and found to have the following properties: \\
(1)~$\hat{W}(n)$ mainly affect a shift in the gauge coupling $\beta$, and have almost no effects in the determination of the phase structure. \\
(2)~$\hat{L}_m (N_t, n)$ for $m \geq 2$ are much smaller than $\hat{L}_1 (N_t, n)$. \\
(3)~$\hat{L} (N_t, n)$ is strongly correlated with the Polyakov loop ${\rm Re} \Omega$ on each configuration, i.e.,
\begin{eqnarray}
\hat{L} (N_t, n) \approx L^0 (N_t, n) c_n {\rm Re} \Omega,
\label{eq:lnapp}
\end{eqnarray}
where $\hat{L} (N_t, n) = \sum_m [\hat{L}_m^+ (N_t, n)+ \hat{L}_m^- (N_t, n)] =  2\sum_m {\rm Re} \hat{L}_m^+ (N_t, n)$, $c_n$ is a proportionality constant, and $L^0(N_t, n)$ is $\hat{L}(N_t, n)$ when all link fields $U_{\mu}(x)$ are set to 1, which is given in Table 2 of Ref.~\cite{Wakabayashi:2021eye}. \\
(4)~The arguments of those complex numbers are approximately identical \cite{Ejiri:2023tdp}, 
\footnote{Note that Eq.~(\ref{eq:argapp}) is not satisfied when ${\rm Re} \hat{L}_1^+(N_t, n)$ becomes a negative number.
In the case of $N_t=6$, the sign of ${\rm Re} \hat{L}_1^+(N_t, n)$ changes to negative number from $n=20$ \cite{Ejiri:2023tdp}.}
\begin{eqnarray}
{\rm Arg} \hat{L}_1^+ (N_t, n) \approx {\rm Arg} \Omega, \hspace{3mm} {\rm and} \hspace{3mm}
2 {\rm Im} \hat{L}_1^+ (N_t, n) \approx L^0 (N_t, n) c_n {\rm Im} \Omega .
\label{eq:argapp}
\end{eqnarray}

These properties of the expansion coefficients allow us to replace bent Polyakov loops with linear Polyakov loops $\Omega$ in a relatively wide region of $\kappa$ where the hopping parameter expansion is valid, i.e.,
\begin{eqnarray}
\sum_{n=N_t}^{\infty} \hat{L}_1^{+} (N_t, n) e^{\mu/T} \kappa^n 
&\approx& \Omega e^{\mu/T} \sum_{n=N_t}^{\infty} L_1^0 (N_t, n) c_n \kappa^n
\equiv 6 \times 2^{N_t} N_t^{-1} \Omega e^{\mu/T} \kappa_{\rm eff}^{N_t} ,
\nonumber \\
\sum_{n=N_t}^{\infty} \hat{L}_1^{-} (N_t, n) e^{-\mu/T} \kappa^n 
&\approx& \Omega^* e^{-\mu/T} \sum_{n=N_t}^{\infty} L_1^0 (N_t, n) c_n \kappa^n
\equiv 6 \times 2^{N_t} N_t^{-1} \Omega^* e^{-\mu/T} \kappa_{\rm eff}^{N_t} .
\end{eqnarray}
This means that the effect of the $\hat{L}_m^{\pm} (N_t, n)$ term containing spatial links can be incorporated into the expansion of $\det M$, which ignores the terms containing spatial links, by shifting $\kappa$ to $\kappa_{\rm eff}$.
We want to investigate the region where $\mu$ is very large.
However, when $\mu$ is large, research is not possible due to the sign problem.
As a trade-off, we make the dynamical quark heavier (reduce $\kappa$) to suppress the sign problem.
In lattice QCD, the theory that approximates it by increasing the quark mass and ignoring the spatial link term is called the heavy dense effective theory of QCD \cite{Blum:1995cb,Aarts:2008rr,Aarts:2017vrv}.
Here we think it is worthwhile to revisit the heavy dense effective theory and discuss QCD at very high densities.
By shifting the hopping parameter $\kappa$ to $\kappa_{\rm eff}$, we consider that the heavy dense effective theory is valid over a wide range of $\kappa$, but for simplicity we will write $\kappa$ instead of $\kappa_{\rm eff}$.

\subsection{Lattice QCD with heavy quarks at high densities}

The partition function of lattice QCD with $N_{\rm f}$ flavors of quarks is
\begin{eqnarray}
{\cal Z} =\int{\cal D} U \det M(U)^{N_{\rm f}} e^{-S_{\rm g} (U)},
\end{eqnarray}
where $U_{\mu}(x)$ is the $SU(3)$ link field, $S_g$ is the action of the gauge field, and 
$M$ is the quark kernel on the lattice.
We adopt the standard Wilson fermion in this work.
The Wilson quark kernel is
\begin{eqnarray}
M_{x,y} &=& \delta_{x,y} - \kappa \sum_{j=1}^3 \{(1-\gamma_j)U_j (x) \delta_{y,x+\hat{j}}+(1+\gamma_j)U_j^\dagger (y) \delta_{y,x-\hat{j}} \} \nonumber \\
&& - \kappa(1-\gamma_4)U_4 (x) e^{\mu a}\delta_{y,x+\hat{4}}-\kappa (1+\gamma_4)U_4^\dagger (y) e^{-\mu a}\delta_{y,x-\hat{4}}, 
\label{eq:wilk}
\end{eqnarray}
where $\kappa$ is the hopping parameter, $\mu$ is the chemical potential, and $a$ is the lattice spacing.
The lattice size is $N_s^3 \times N_t$.
When the quark mass is heavy and the chemical potential is large, i.e. $\kappa \rightarrow 0, e^{\mu a} \rightarrow \infty$, the quark kernel can be simplified to
\begin{eqnarray}
M_{x,y}=\delta_{x,y} - \kappa(1-\gamma_4)U_4 (x) e^{\mu a} \delta_{y,x+\hat{4}}.
\label{eq:wilkap}
\end{eqnarray}
The temperature $T$ is defined as the inverse of the length in the time direction over which the periodic boundary conditions are imposed, and is $T=1/(N_t a)$.
Since there is no link field in the spatial direction, the quark determinant can be expressed as a product of the determinants at each spatial point $\vec{x}$.

Taking into account the antiperiodic boundary condition of the fermion field, the quark determinant can be written as follows:
\begin{eqnarray}
\mathrm{det}M &=& \prod_{\vec{x}}
  \begin{vmatrix}
  I & -\kappa(1-\gamma_4)U_4 (\vec{x}) e^{\mu a} & 0 & \ldots & 0 \\
  0 & I & -\kappa(1-\gamma_4)U_4 (\vec{x}+\hat{4}) e^{\mu a} & \ldots & 0 \\
  0 & 0 & I & \ldots & 0 \\
  \vdots & \vdots & \vdots & \ddots  & \vdots \\
  \kappa(1-\gamma_4)U_4 (\vec{x}-\hat{4}) e^{\mu a} & 0 & 0 & \ldots & I
  \end{vmatrix} \nonumber \\
&=& \prod_{\vec{x}} \mathrm{det} \left[ I+(\kappa e^{\mu a})^{N_t}(1-\gamma_4)^{N_t} \prod_{i=0}^{N_t-1} U_4 (\vec{x}+i\hat{4}) \right] \nonumber \\
&=& \prod_{\vec{x}} \mathrm{det} \left[ I+2^{N_t-1} \kappa^{N_t}e^{\mu/T}(1-\gamma_4) \prod_{i=0}^{N_t-1} U_4 (\vec{x}+i\hat{4}) \right] ,
\end{eqnarray}
where $I$ is the $(3 \times 3) \otimes (4 \times 4)$ identity matrix.
Due to the periodic boundary conditions, $U_4 (\vec{x}-\hat{4})=U_4 (\vec{x}+(N_t-1)\hat{4})$.
Here we introduce the parameter $C=(2\kappa)^{N_t} e^{\mu/T}$.
Diagonalizing $\prod_{i=0}^{N_t-1} U_4 (\vec{x}+i\hat{4})$ by a unitary matrix $V$ and expressing the quark determinant in terms of the eigenvalues $\lambda_i (\vec{x})$ $(i=1,2,3)$, we get
\begin{eqnarray}
\det M &=& \prod_{\vec{x}} \mathrm{det} \left[ I+\frac{C}{2}(1-\gamma_4)V \prod_{i=0}^{N_t-1} U_4 (\vec{x}+i\hat{4}) V^\dagger \right] \nonumber \\ 
&=& \prod_{\vec{x}} \left[ (1+C\lambda_1(\vec{x}))(1+C\lambda_2(\vec{x}))(1+C\lambda_3(\vec{x})) \right]^2.
\end{eqnarray}
Since $\prod_{i=0}^{N_t-1} U_4 (\vec{x}+i\hat{4})$ is an $SU(3)$ matrix, 
the eigenvalues satisfy the following equations:
\begin{eqnarray}
&&\lambda_1 \lambda_2 \lambda_3 = 1, \hspace{10mm}
\lambda_1 \lambda_1^*=\lambda_2 \lambda_2^*=\lambda_3 \lambda_3^*=1,  \\
&&\lambda_1(\vec{x}) +\lambda_2(\vec{x}) +\lambda_3(\vec{x})
=\mathrm{tr} \prod_{i=0}^{N_t-1} U_4 (\vec{x}+i\hat{4}) = 3 \Omega^{\rm (loc)} (\vec{x}) ,
\end{eqnarray}
where $\Omega^{\rm (loc)} (\vec{x})$ is the local Polyakov loop at each spatial point $\vec{x}$ defined as
\begin{eqnarray}
\Omega^{\rm (loc)} (\vec{x}) = \frac{1}{3} \mathrm{tr}\prod_{i=0}^{N_t-1} U_4 (\vec{x}+i\hat{4}).
\end{eqnarray}
Using these equations, the quark determinant becomes
\begin{eqnarray}
\mathrm{det}M = \prod_{\vec{x}} \left\{1+3C \Omega^{\rm (loc)} (\vec{x}) + 3C^2 \left( \Omega^{\rm (loc)} (\vec{x}) \right)^* + C^3 \right\}^2.
\end{eqnarray}

Let us suppose that a critical point is found and call it $C_c$.
Since this effective theory is controlled only by $C=(2\kappa)^{N_t} e^{\mu/T}$, when $\kappa$ is changed, the critical chemical potential $\mu_c$ changes as 
$\mu_c/T = \ln C_c - N_t \ln(2 \kappa)$.
When $\kappa=0$ (quenched limit), $\mu_c$ is infinite.
Increasing $\kappa$ from 0 decreases the critical chemical potential, approaching the critical $\kappa$ at $\mu =0$ in the heavy-quark region.

\subsection{High-density/low-density duality}

The quark determinant of the heavy dense effective theory can be rewritten as
\begin{eqnarray}
\mathrm{det}M = C^{6 N_s^3} \prod_{\vec{x}} \left\{1+3C^{-1} \left( \Omega^{\rm (loc)} (\vec{x}) \right)^* + 3C^{-2} \Omega^{\rm (loc)} (\vec{x}) + C^{-3} \right\}^2.
\label{eq:omegaloc}
\end{eqnarray}
$C$ is a constant parameter of the theory, and the overall constant does not affect the calculation of the path integral.
This indicates that the effective theory is invariant under transformations that transform $C$ into $C^{-1}$ and simultaneously swap $\Omega^{\rm (loc)}$ and $\Omega^{\rm (loc) *}$ \cite{Blum:1995cb}.
Since QCD has a symmetry under the transformation that makes the link variable $U_{\mu}(x)$ Hermitian conjugate, this symmetry of swapping $C$ and $C^{-1}$ is a symmetry between low density and high density if we fix the mass to be heavy.
When $C$ is infinite, the $\Omega^{\rm (loc)}$ and $\Omega^{\rm (loc)*}$ terms become ineffective and the quark determinant becomes a constant, which is the same as in quenched QCD with $C=0$.

If the critical point is found at $C_c$, then due to this symmetry, there will be two critical chemical potentials $\mu_c$, with $\mu_c/T = \pm \ln C_c - N_t \ln(2 \kappa)$.
Therefore, decreasing the quark mass (increasing $\kappa$) decreases both the large and small critical chemical potentials.
The critical chemical potential in the high-density region may be related to the critical point in the light-quark, low-density region.

This high-density/low-density duality is not a consequence of the usual particle-antiparticle symmetry of QCD, but rather a symmetry that emerges from the disappearance of the $U_4^{\dagger} (x)$ terms at high densities in the process from Eq.~(\ref{eq:wilk}) to Eq.~(\ref{eq:wilkap}).
In other words, it is a consequence of particle-hole symmetry.

\subsection{Effective three-dimensional spin model}
\label{sec:eff3dmodel}

We consider the three-dimensional three-state Potts model \cite{Wu:1982ra} as an effective model of heavy dense QCD.
These two models are both three-dimensional space theories and have the same broken symmetry, $Z_3$, so they are thought to have the same phase structure.
Similar to the early days of lattice QCD when the effect of dynamical quarks on the finite temperature phase transition was discussed in comparison with the Potts model \cite{DeGrand:1983fk}, we replace the Polyakov loop with a $Z_3$ spin variable.
The integral measure and gauge field action are replaced by the spin variable, $s(\vec{x})$,
\begin{eqnarray}
\int {\cal D} U \ \rightarrow \ \sum_{s(\vec{x})} , \qquad
S_{\rm g} (U) \ \rightarrow \ -\beta\sum_{\vec{x}} \sum_{i=1}^3
\mathrm{Re} \left[ s (\vec{x}) s^* (\vec{x}+\hat{i}) \right]
\end{eqnarray}
Here, $s (\vec{x})$ can take the following three states:
\begin{eqnarray}
s_1=1, \quad s_2=e^{2\pi i/3}, \quad s_3=e^{-2\pi i/3}.
\end{eqnarray}
In the absence of an external magnetic field, this model has $Z_3$ symmetry.

We show that the equivalent of the quark determinant is the complex external magnetic field term.
Replacing Polyakov loops $3 \Omega^{\rm (loc)}  (\vec{x})$ with spins $s  (\vec{x})$,
\begin{eqnarray}
(\mathrm{det}M)^{N_{\rm f}} \rightarrow \prod_{\vec{x}}(1+C s (\vec{x}) + C^2 s^* (\vec{x}) + C^3 )^{2N_{\rm f}}
\equiv \prod_{\vec{x}} F (\vec{x}).
\end{eqnarray}
If $s (\vec{x})=s_1$,
\begin{eqnarray}
F (\vec{x}) = (1 + C + C^2 + C^3)^{2N_{\rm f}} \equiv e^{A_1},
\end{eqnarray}
if $s (\vec{x})=s_2$,
\begin{eqnarray}
F (\vec{x}) = \left[ 1+C(-\frac{1}{2}+\frac{\sqrt{3}}{2}i) + C^2(-\frac{1}{2}-\frac{\sqrt{3}}{2}i) + C^3 \right]^{2N_{\rm f}} \equiv e^{A_2 +i\theta},
\end{eqnarray}
if $s (\vec{x})=s_3$, 
\begin{eqnarray}
F (\vec{x}) = \left[ 1+C(-\frac{1}{2}-\frac{\sqrt{3}}{2}i) + C^2(-\frac{1}{2}+\frac{\sqrt{3}}{2}i) + C^3 \right]^{2N_{\rm f}} \equiv e^{A_2 - i\theta}.
\end{eqnarray}
Assuming the number of spins taking state $s_i$ is $N_i$ and the total number of spins is $N_{\rm site}=N_1+N_2+N_3$, from the above equations, $\prod_{\vec{x}} F (\vec{x})$ is given by the following equation:
\begin{eqnarray}
\prod_{\vec{x}} F (\vec{x}) &=& \mathrm{exp} \left[ {N_1A_1+N_2(A_2 + i\theta)+N_3(A_2 - i\theta)} \right] \nonumber \\
&=& \mathrm{exp} \left[ (A_1-A_2)N_1 + A_2N_{\rm site}+i\theta(N_2-N_3) \right].
\end{eqnarray}
Therefore, the partition function is
\begin{eqnarray}
{\cal Z} = \sum_{s(\vec{x})} \exp \left[
\beta\sum_{\vec{x}}\sum_{i=1}^3 \mathrm{Re} \left[ s (\vec{x}) s^* (\vec{x}+\hat{i}) \right] + (A_1-A_2)N_1 + i\theta(N_2-N_3) + A_2N_{\rm site}
\right] . \hspace{5mm}
\end{eqnarray}
Furthermore, from the relationship between $N_i$ and $s_i$,
\begin{eqnarray}
\sum_{\vec{x}} \mathrm{Re} \left[ s (\vec{x}) \right] = N_1 - \frac{1}{2}(N_2 + N_3) , &\hspace{5mm}&
\sum_{\vec{x}} \mathrm{Im} \left[ s (\vec{x}) \right] = \frac{\sqrt{3}}{2} (N_2-N_3), \nonumber \\
N_1 = \frac{2}{3}\sum_{\vec{x}} \mathrm{Re} \left[ s (\vec{x}) \right] + \frac{1}{3} N_{\rm site} , &\hspace{5mm}&
N_2-N_3 = \frac{2}{\sqrt{3}}\sum_{\vec{x}} \mathrm{Im} \left[ s (\vec{x}) \right] ,
\end{eqnarray}
we can rewrite ${\cal Z}$ as
\begin{eqnarray}
{\cal Z} = \sum_{s(\vec{x})} \exp \left[
\beta\sum_{\vec{x}} \sum_{i=1}^3 \mathrm{Re} \left[ s (\vec{x}) s^* (\vec{x}+\hat{i}) \right]
+h\sum_{\vec{x}} \mathrm{Re} [s (\vec{x})]
+iq\sum_{\vec{x}} \mathrm{Im} [s (\vec{x})]
\right] ,
\label{eq:Zpott}
\end{eqnarray}
excluding the overall constant.
This is the partition function of the three-dimensional three-state Potts model, but with terms for real and imaginary external fields.
When corresponding to heavy-quark high-density QCD, the parameters $h$ and $q$ are real numbers and can be expressed as functions of $C$ as follows:
\begin{eqnarray}
h &=& \frac{4}{3}N_{\rm f} \mathrm{ln}(1+C+C^2+C^3)
- \frac{2}{3}N_{\rm f} \mathrm{ln} \left[ \left(1-\frac{1}{2}C -\frac{1}{2}C^2 +C^3\right)^2 + \frac{3}{4}(C-C^2)^2 \right], 
\label{eq:pottsh}
\\
q &=& \frac{4}{\sqrt{3}}N_{\rm f} \ \mathrm{arctan}\left[ 
\frac{\frac{\sqrt{3}}{2}(C-C^2)}{1-\frac{C}{2}-\frac{C^2}{2}+C^3}
\right].
\label{eq:pottsq}
\end{eqnarray}

\begin{figure}[tb]
\begin{minipage}{0.47\hsize}
\begin{center}
\vspace{0mm}
\includegraphics[width=7.7cm]{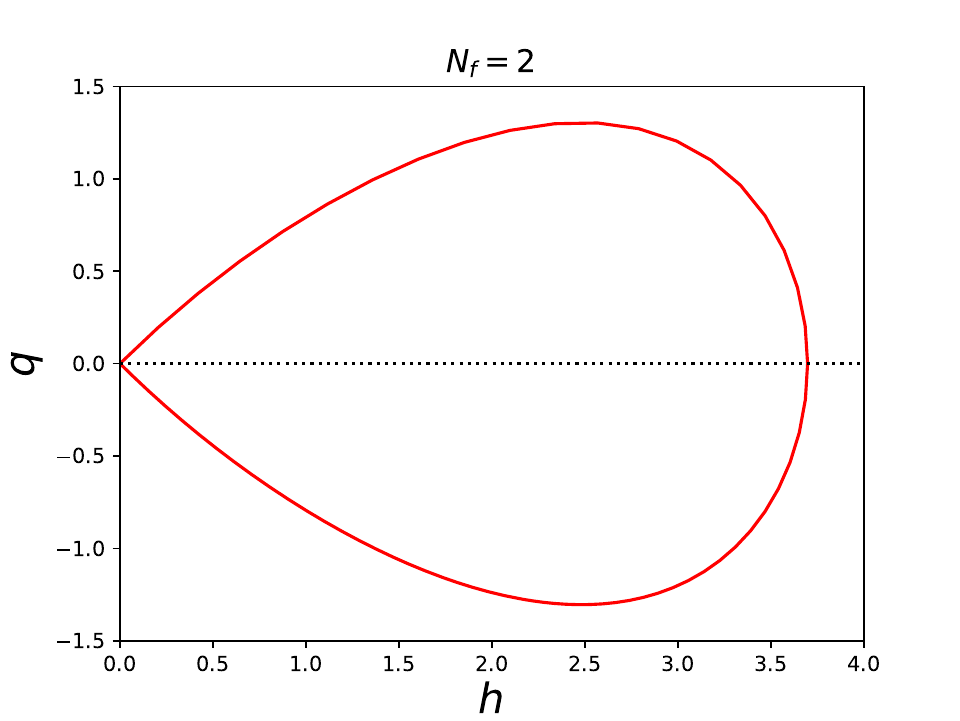}
\vspace{-6mm}
\end{center}
\caption{The corresponding parameters $(h,q)$ of the spin model when changing $C$ in heavy dense QCD for $N_{\rm f}=2$.}
\label{fig:hq2}
\end{minipage}
\hspace{2mm}
\begin{minipage}{0.47\hsize}
\begin{center}
\vspace{0mm}
\includegraphics[width=7.7cm]{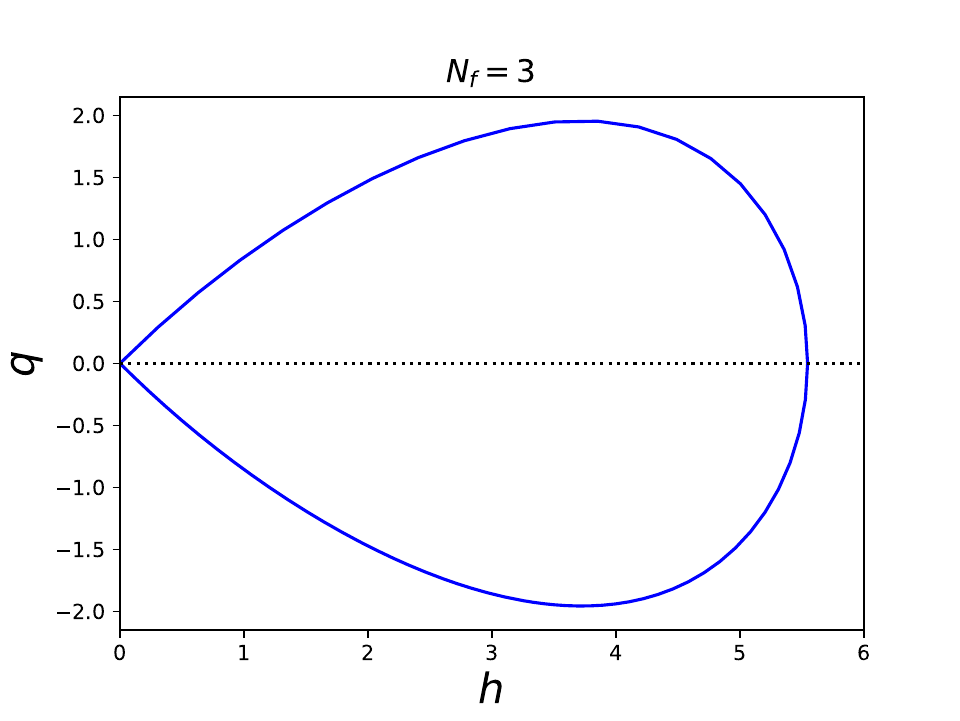}
\vspace{-6mm}
\end{center}
\caption{The same figure as Fig.~\ref{fig:hq2} for $N_{\rm f}=3$. }
\label{fig:hq3}
\end{minipage}
\end{figure}

In Figs.~{\ref{fig:hq2} and \ref{fig:hq3}, we plot the trajectory in the parameter space of $(h, q)$ as we vary $C$ from zero to infinity with the red and blue curves.
Figures~{\ref{fig:hq2} and \ref{fig:hq3} show the results for two flavors and three flavors, respectively.
When $C=0$ and $C=\infty$, $(h, q)=(0, 0)$.
For $C<1$, $q>0$, for $C=1$, $q=0$, and for $C>1$, $q<0$.
As explained in the previous section, due to the symmetry of $C \leftrightarrow C^{-1}$, this trajectory is symmetric under the interchange of $q \leftrightarrow -q$.
This Potts model has $q \leftrightarrow -q$ symmetry.
When $C$ is small near $(h,q)=(0,0)$, $h \approx 2 N_{\rm f} C$ and $q \approx 2 N_{\rm f} C$, so the slope of the trajectory at $(h,q)=(0,0)$ is 1, i.e., $dq/dh=1$.
As $C$ varies from zero to infinity, this trajectory returns to the same point with a finite length, making it easy to explore the entire parameter space corresponding to the heavy dense QCD.

\subsection{Histogram of magnetization}
\label{sec:histogram}

\begin{figure}[tb]
\begin{center}
\vspace{0mm}
\includegraphics[width=7.0cm]{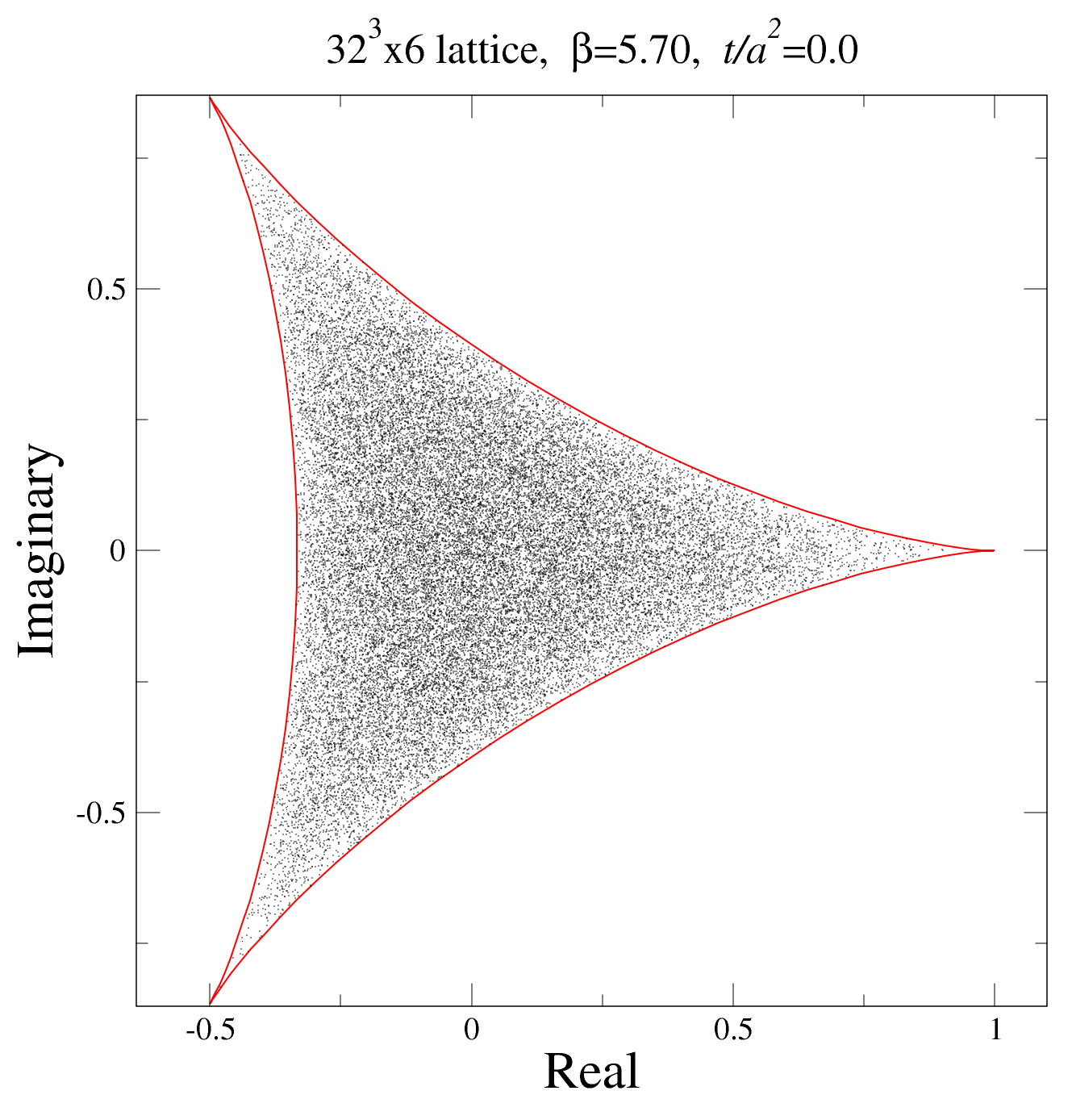}
\hspace{2mm}
\includegraphics[width=7.0cm]{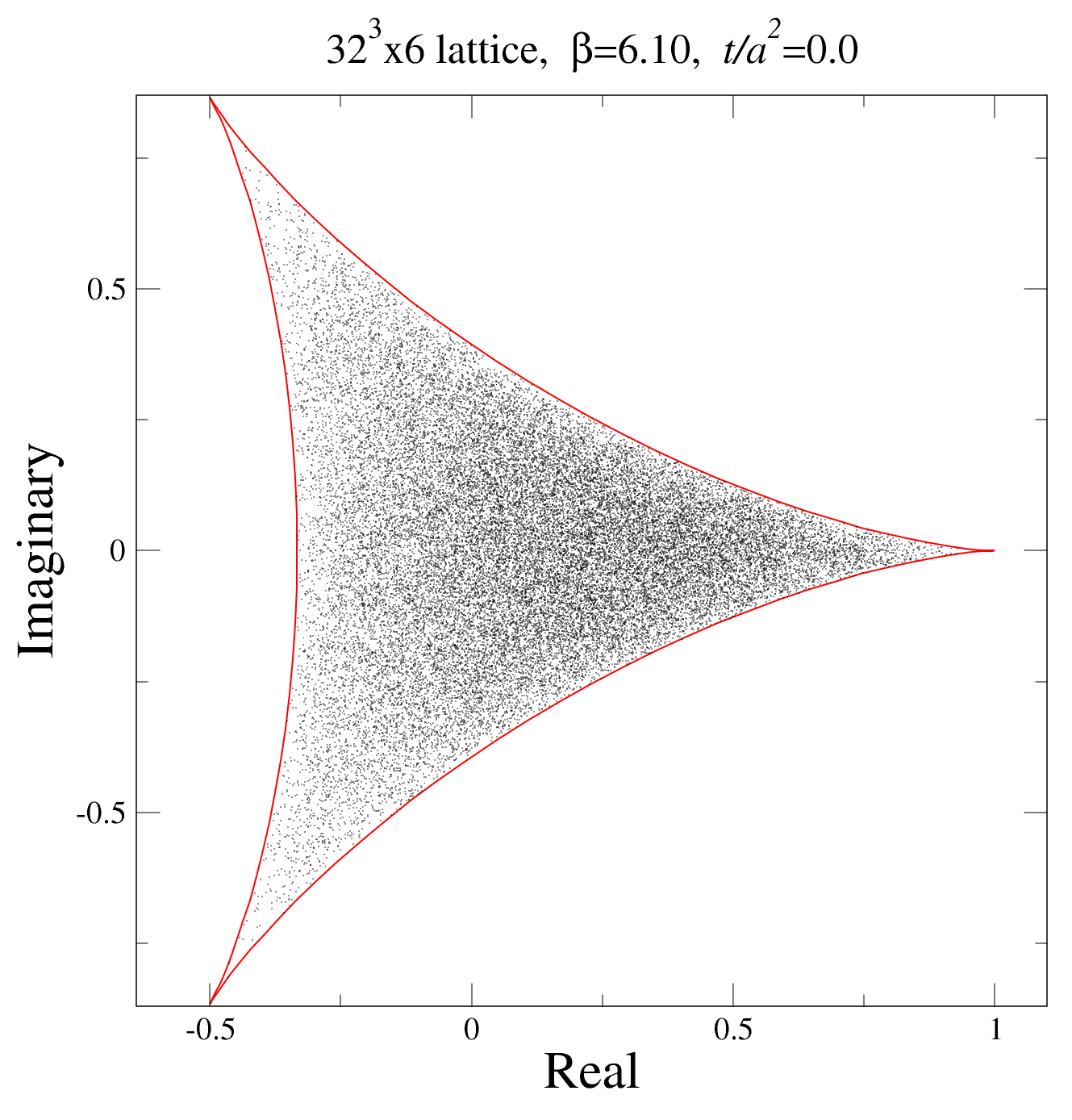}
\vspace{-5mm}
\end{center}
\caption{Distribution of the local Polyakov loop at each point in one configuration of quenched QCD.
The left panel shows the distribution for the symmetric phase $(\beta =5.70)$, and the right panel shows the distribution for the broken phase $\beta = 6.10)$.}
\label{fig:pl3hist}
\end{figure}

In the previous section, we replaced the Polyakov loop with a $Z_3$ spin, and here we discuss its validity.
Since $\Omega^{\rm (loc)} (\vec{x})$ of Eq.~(\ref{eq:omegaloc}) is the trace of an $SU(3)$ matrix, its values are distributed within the distorted triangle in the complex plane  enclosed by the red line in Fig.~\ref{fig:pl3hist} \cite{Ejiri:2022jai}.
The horizontal axis is the real part of  $\Omega^{\rm (loc)}$, and the vertical axis is the imaginary part of  $\Omega^{\rm (loc)}$.
On the other hand, $Z_3$ spin is a variable that takes on the values 
$s (\vec{x}) = 1$, $e^{2\pi i/3}$, and $e^{-2\pi i/3}$.
Although these variables at each point are quite dissimilar, the probability distributions of their spatial averages in the complex plane are very similar.
Figure \ref{fig:shist} shows the probability distribution of the spatial average of spin values, i.e., the magnetization, calculated by a simulation of the three-state Potts model on a $40^3$ lattice.
The number of configurations is 200000.
The value of $\beta$ is adjusted to the phase transition point for each $h$ by the reweighting method.
The transition point is considered to be the point where the susceptibility is maximized, 
and the peak position is $\beta =0.366967(6)$ for $h=0.0$, $\beta =0.366304(4)$ for $h=0.00051$, 
and $\beta =0.366178(4)$ for $h=0.0006$ for $N_{\rm site}=40^3$.
The top left panel shows the results calculated with $h$ and $q$ set to zero in the Boltzmann weight of Eq.~(\ref{eq:Zpott}).\footnote{
Since the system has $Z_3$ symmetry, we performed a $Z_3$ rotation to symmetrize the distribution function.}
This is a typical probability distribution (histogram) of a first-order phase transition.
In the confinement phase, the Polyakov loop values are distributed around the origin, while in the deconfinement phase, they are distributed around three $Z_3$ symmetric points.
At the first-order phase transition, these two phases coexist, and the Polyakov loop values are distributed around these four points.

\begin{figure}[tb]
\begin{center}
\vspace{0mm}
\includegraphics[width=8.0cm]{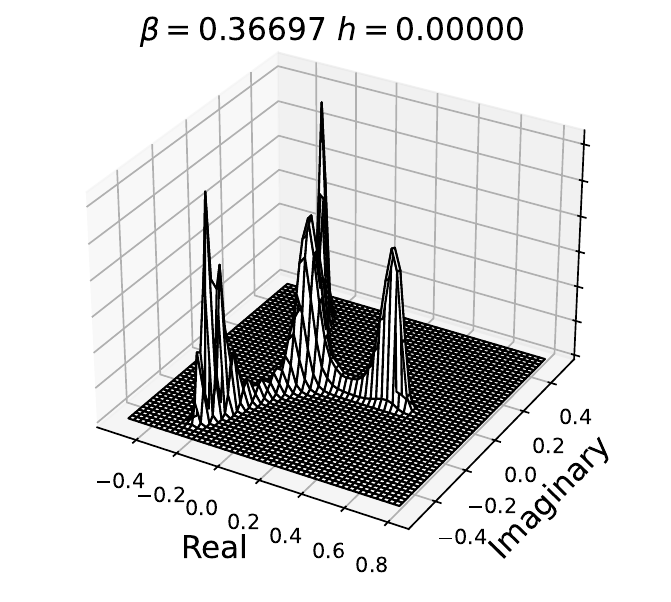}
\hspace{-2mm}
\includegraphics[width=8.0cm]{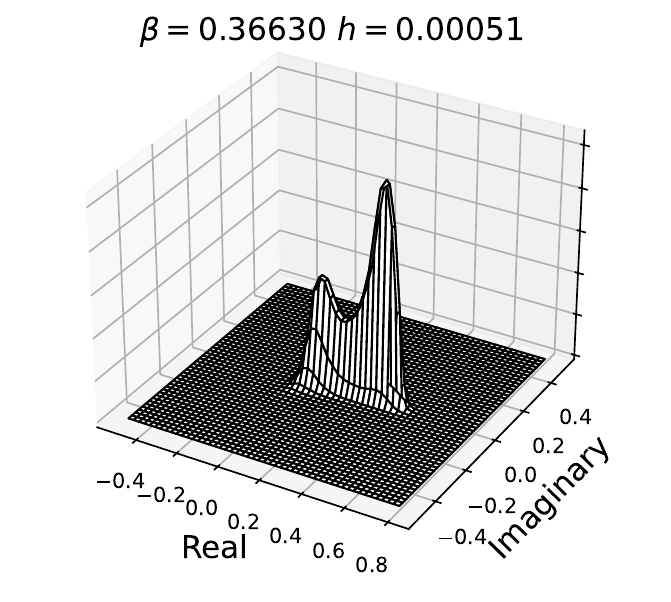}
\\
\vspace{-1mm}
\includegraphics[width=8.0cm]{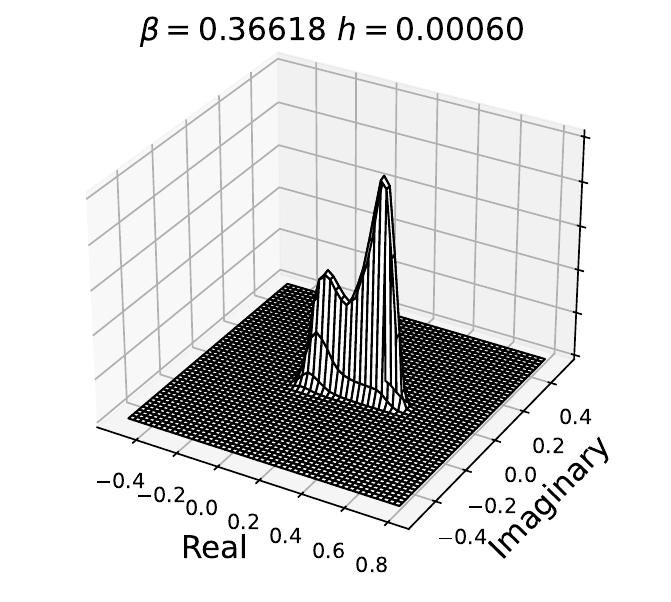}
\vspace{-5mm}
\end{center}
\caption{Probability distribution of magnetization in the complex plane for the three-state Potts model with $q=0$.
The values of $\beta$ and $h$ are shown above the figure.
}
\label{fig:shist}
\end{figure}

In the standard Potts model with $q=0$, increasing the value of $h$ causes the phase transition to become a crossover \cite{DeGrand:1983fk,Karsch:2000xv}. 
The top right panel shows the result for $h$ slightly below the critical point $(h=0.00051)$, and the bottom panel shows the result in the crossover region just above the critical point $(h=0.0006)$ with $q=0$.
These histograms in the complex plane are very similar to the histograms of the Polyakov loop in the heavy-quark region of QCD.
See Fig.~6 in Ref.~\cite{Saito:2013vja}.
The susceptibility and higher-order cumulants, which are used to determine the universality class at the critical point, are uniquely determined by the shape of the histogram of the order parameter.
Therefore, how the shape of the histogram changes before and after the phase transition determines the fundamental properties of that phase transition.

\begin{figure}[tb]
\begin{center}
\vspace{0mm}
\includegraphics[width=7.0cm]{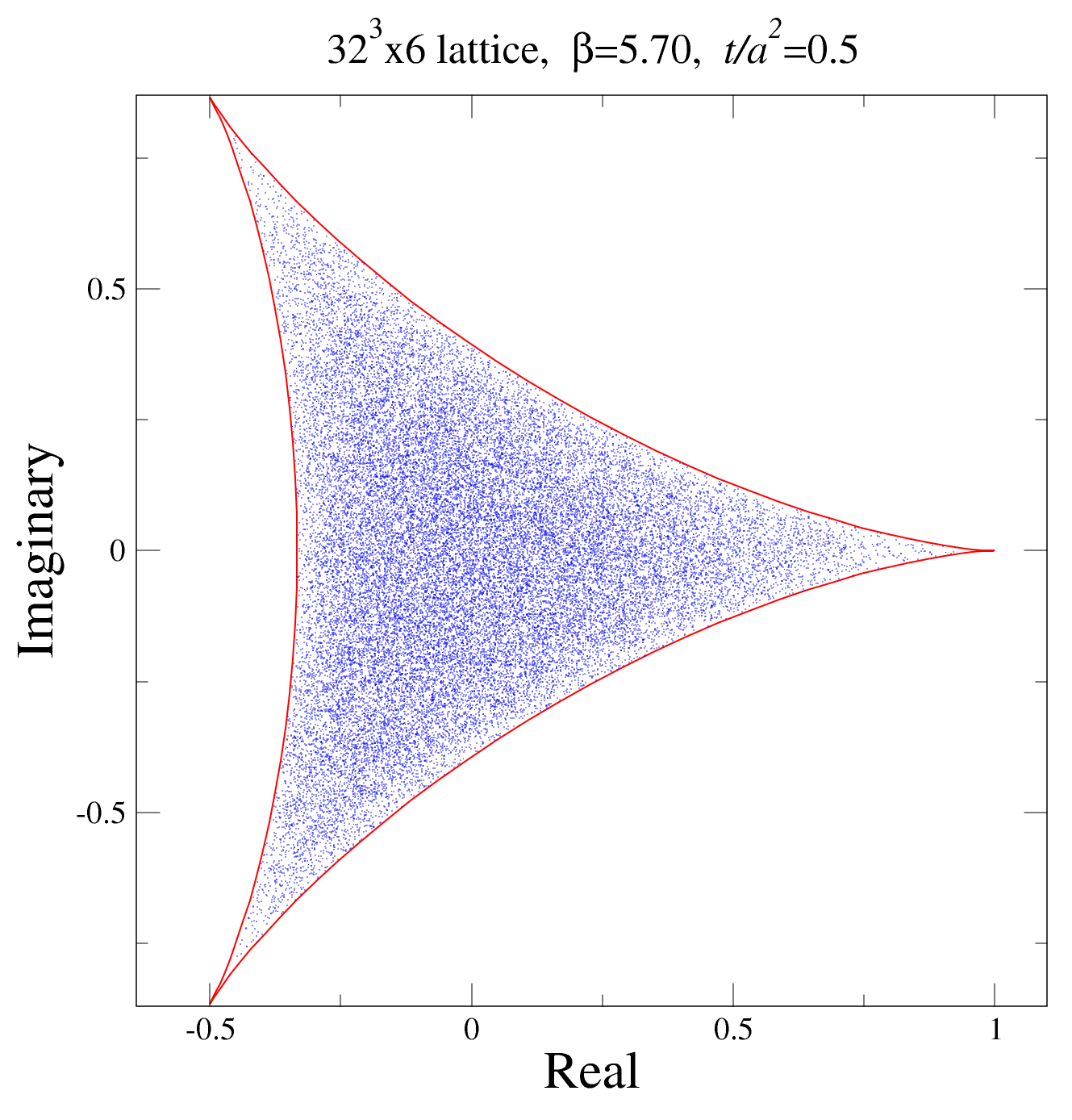}
\hspace{2mm}
\includegraphics[width=7.0cm]{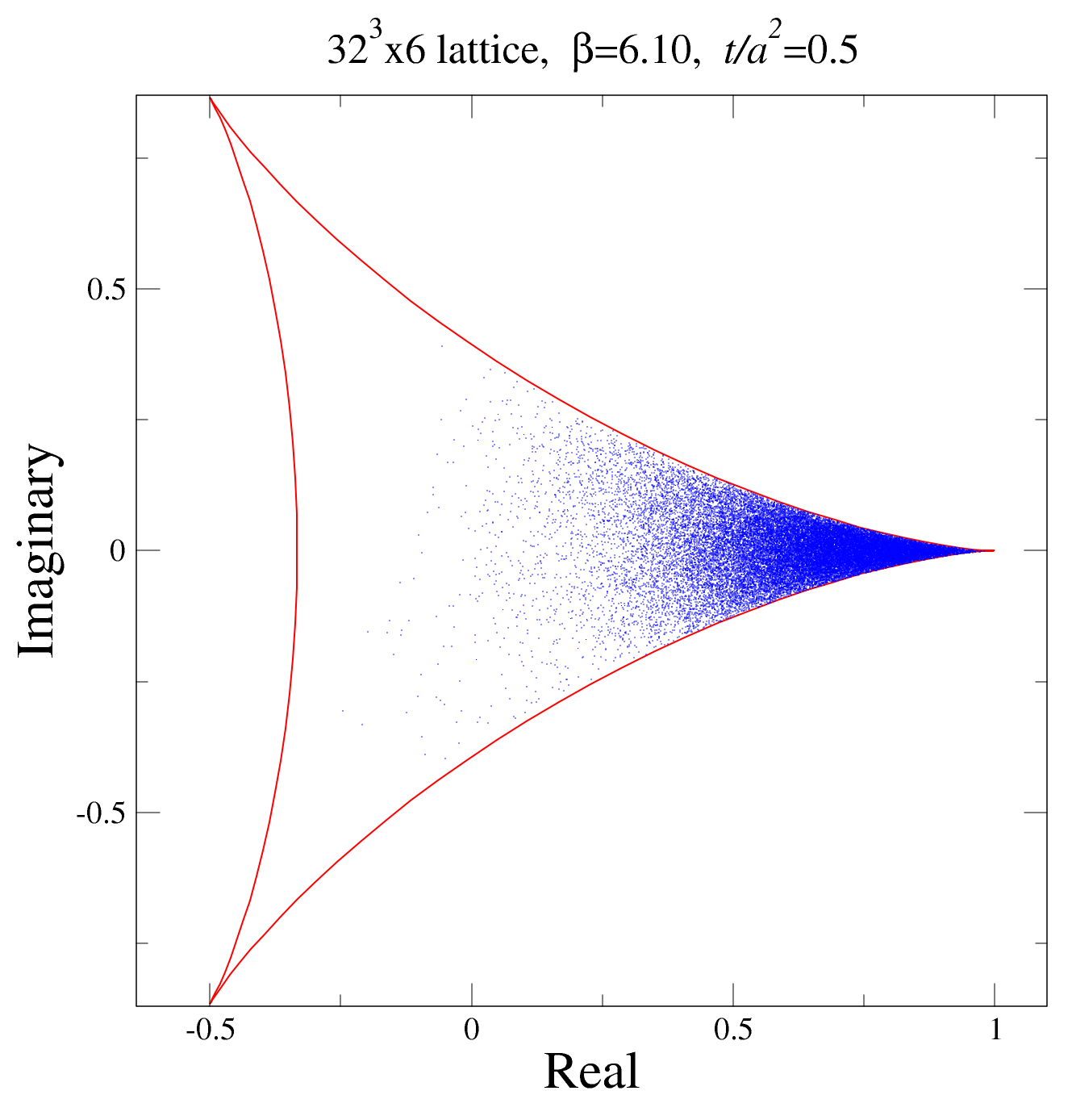}
\vspace{-3mm}
\end{center}
\caption{Distribution of the local Polyakov loop at each point in one configuration of quenched QCD after coarse graining by the gradient flow at $t/a^2=0.5$ for $\beta =5.70$ (left) and $6.10$ (right).}
\label{fig:pl3gradhist}
\end{figure}

To discuss spontaneous symmetry breaking in the absence of an external magnetic field $h$, it is more appropriate to focus on the local Polyakov loops or spin values before averaging.
In Fig.~\ref{fig:pl3hist}, we plot the values of the Polyakov loop $\Omega^{\rm (loc)} (\vec{x})$ in the complex plane at each point for a typical gauge field configuration.
The left panel shows the result for the confinement phase $(\beta =5.70)$ calculated by quenched QCD simulations $(h=q=0)$ on a $32^3 \times 6$ lattice and the right panel shows the result for the deconfinement phase $(\beta =6.10)$.
In the confinement phase, the distribution shows $Z_3$ symmetry, whereas in the deconfinement phase, as shown in the right figure, the distribution is slightly skewed toward the right.
The slight asymmetry in the distribution shown in the right figure is barely noticeable. 
However, when the gauge field configuration is coarse-grained using gradient flow, it becomes clear that the symmetry is broken.
In the gradient flow method of QCD proposed in Ref.~\cite{Narayanan:2006rf,Luscher:2009eq,Luscher:2010iy}, the ``flowed'' gauge field $B_{\mu}^a(t,x)$ at flow time $t$ is obtained by solving the flow equation, 
\begin{equation}
   \frac{\partial}{\partial t} B_\mu^a (t,x)=D_\nu G_{\nu\mu}^a(t,x) \equiv
\partial_\nu G_{\nu\mu}^a(t,x)+f^{abc}B_\nu^b(t,x)G_{\nu\mu}^c(t,x) 
\label{eq:flow}
\end{equation}
for quenched QCD with the initial condition $B_{\mu}^a(0,x)=A_{\mu}^a(x)$, where $G_{\mu\nu}^a(t,x)$
is the flowed field strength given from $B_\mu^a(t,x)$.
Because Eq.~(\ref{eq:flow}) is a kind of diffusion equation, we can regard $B_{\mu}^a(t,x)$ as a smeared field of the original gauge field $A_{\mu}^a(x)$ over a physical range of $\sqrt{8t}$ in four-dimensional space.
We solved the discretized version of the flow equations on lattices.
Figure \ref{fig:pl3gradhist} shows the distribution of the Polyakov loops at each point $\Omega^{\rm (loc)} (\vec{x})$, obtained by coarse graining the same configuration as in Fig.~\ref{fig:pl3hist} using the gradient flow with $t/a^2=0.5$.
The distribution in the confinement phase shown on the left is symmetric, whereas the distribution in the deconfinement phase on the right is clearly asymmetric.

On the other hand, since spin can only take on one of three values, symmetry breaking occurs when the number of spins that take on these three values becomes asymmetric.
Similar to the case of the Polyakov loop, when the spin is coarse-grained using a diffusion equation, 
\begin{eqnarray}
\frac{\partial}{\partial t} \tilde{s} (\vec{x}, t) = D \vec{\nabla}^2 \tilde{s} (\vec{x}, t),
\end{eqnarray}
where $D$ represents the diffusion coefficient.
The values of the coarse-grained spin $\tilde{s} (\vec{x},t)$ will be distributed within the triangle shown by the red line in Figs.~\ref{fig:potdifhist1} and \ref{fig:potdifhist2}.
The diffusion equation is defined on the lattice with the lattice spacing $a$ as follows:
\begin{eqnarray}
\tilde{s} (\vec{x}, t+ \Delta t) = (1-6 \tilde{D}) \tilde{s} (\vec{x}, t) 
+ \tilde{D} \sum_{i=1}^3 \left[ \tilde{s} (\vec{x} + \hat{i}, t) + \tilde{s} (\vec{x} - \hat{i}, t) \right] ,
\end{eqnarray}
where $\tilde{D} = D \Delta t/a^2$.
The initial condition is $\tilde{s} (\vec{x}, 0) = s (\vec{x})$.
We plot the distribution of coarse-grained spins at each point of typical configurations in the symmetric and broken phases.
The results for the symmetric phase at $\beta=0.30$ are shown in Fig.~\ref{fig:potdifhist1}, and the results for the broken phase at $\beta=0.40$ are shown in Fig.~\ref{fig:potdifhist2}. 
The coarse-graining time is $Dt/a^2 =0.1$ (left panel) and $0.5$ (right panel).
We adopted a value of the diffusion constant as $D \Delta t/a^2 =0.01$.
The lattice size is $32^3$.
In both the symmetric phase and the broken symmetry phase, the results for $Dt/a^2 =0.5$ are very similar to the distribution of the Polyakov loop at each point $\Omega^{\rm (loc)} (\vec{x})$.
In addition, since $\sum_{\vec{x}} \tilde{s} (\vec{x}, t+ \Delta t) =\sum_{\vec{x}} \tilde{s} (\vec{x}, t) =\sum_{\vec{x}} s(\vec{x})$, the spatial average of the spin values remains unchanged even after the coarse graining.
That is, the value of the magnetization and the external magnetic field terms in the Hamiltonian do not change under the coarse graining.
Considering these arguments, it becomes clear that the seemingly simplistic approach of replacing the Polyakov loop with a $Z_3$ spin has a valid basis.

\begin{figure}[tb]
\begin{center}
\vspace{0mm}
\includegraphics[width=7.0cm]{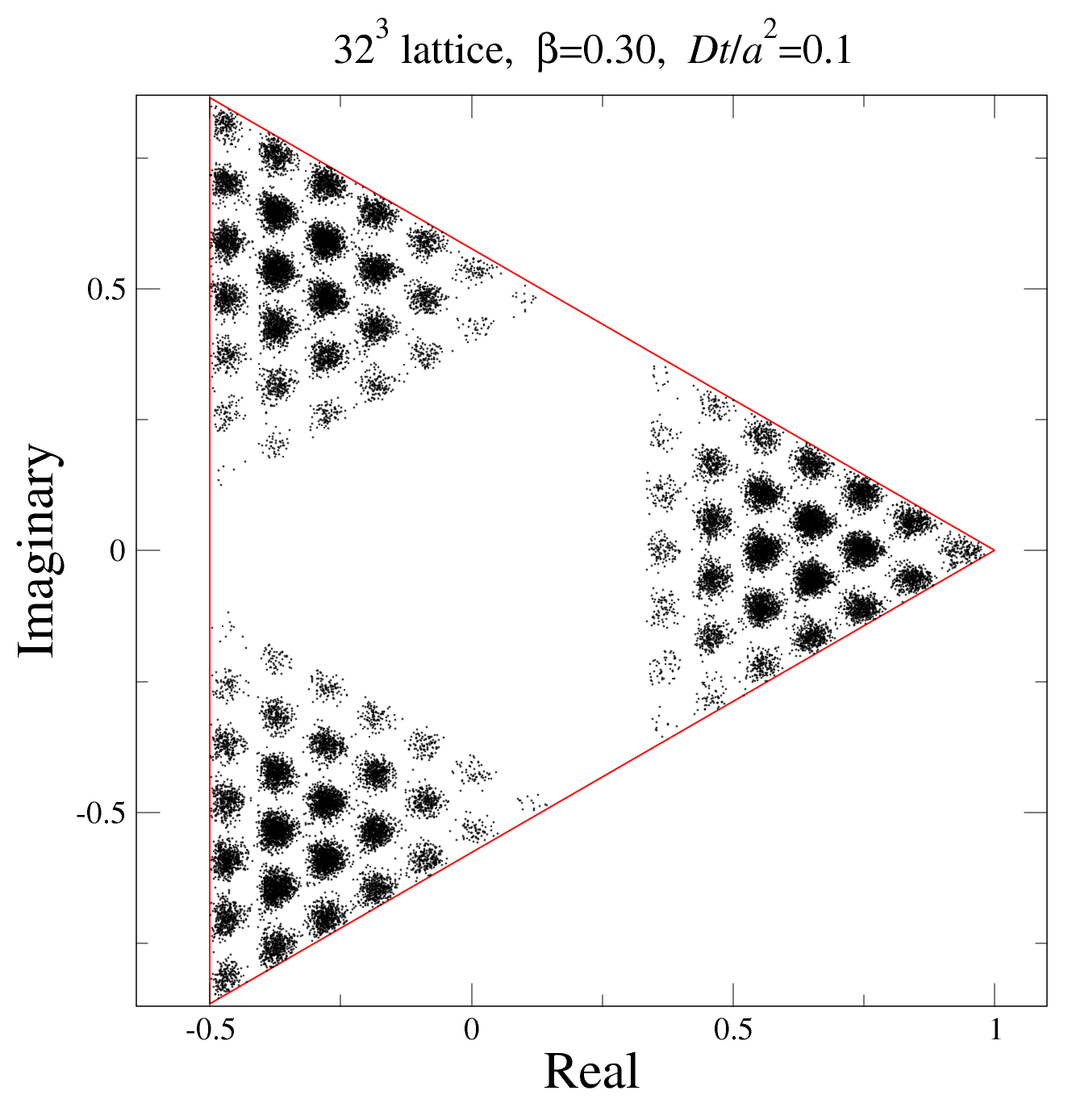}
\hspace{2mm}
\includegraphics[width=7.0cm]{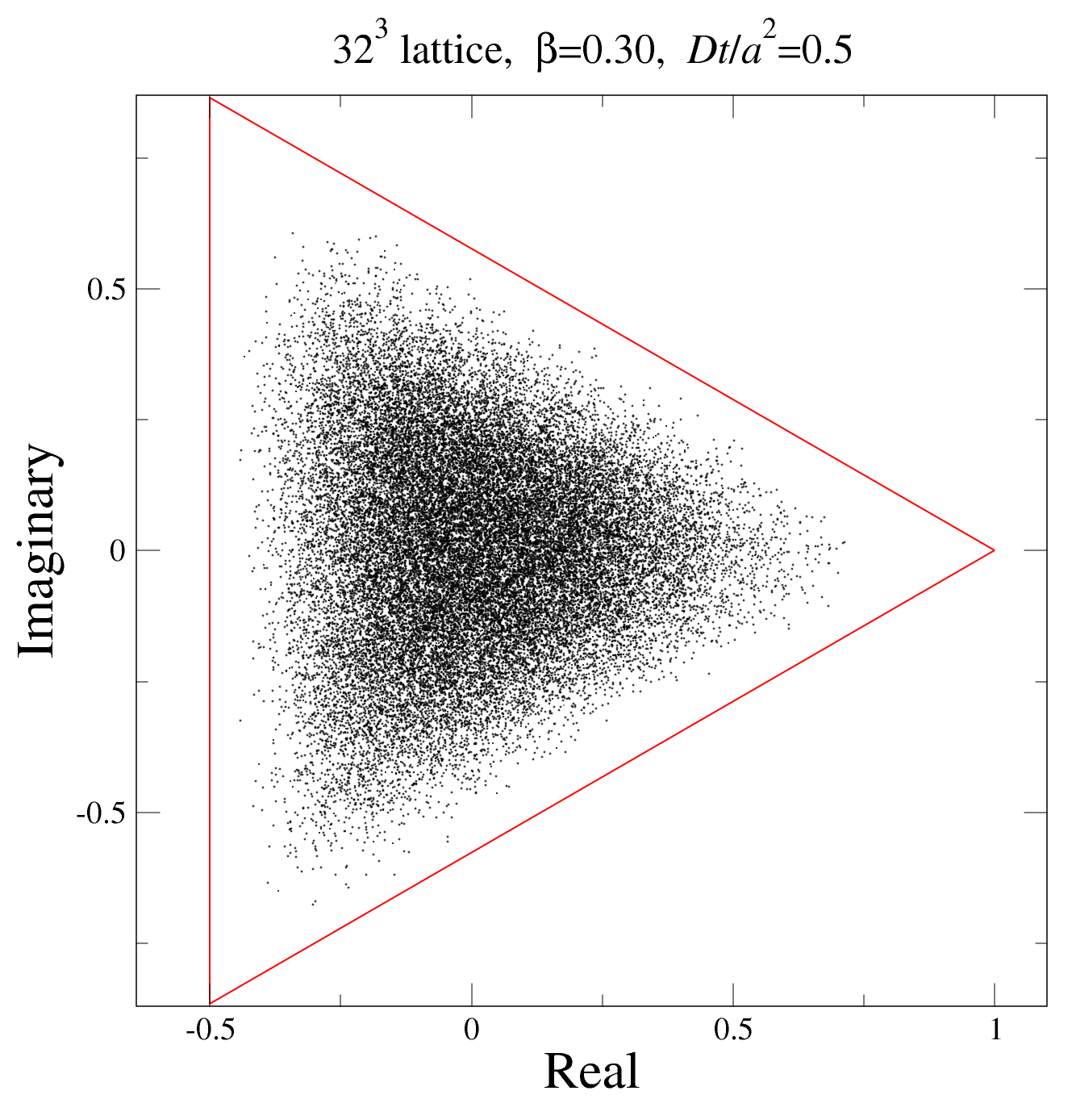}
\vspace{-3mm}
\end{center}
\caption{Distribution of spins $\tilde{s}(\vec{x}, t)$ at each point in one configuration in the symmetric phase $(\beta=0.30)$ of the three-state Potts model after coarse graining with the diffusion equation. 
The left panel is $Dt/a^2 =0.1$, and the right panel is $Dt/a^2 =0.5$.}
\label{fig:potdifhist1}
\end{figure}

\begin{figure}[tb]
\begin{center}
\vspace{0mm}
\includegraphics[width=7.0cm]{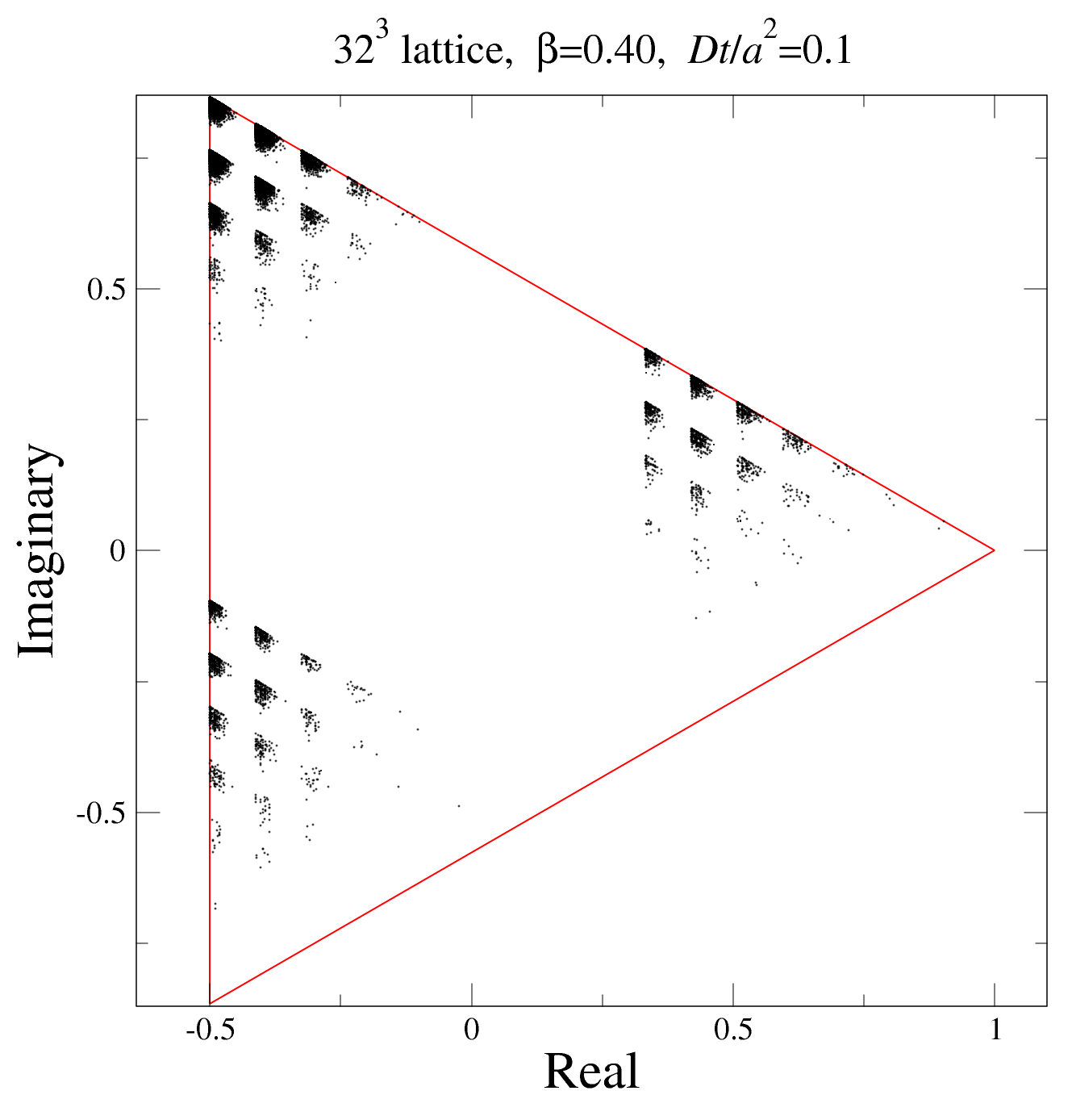}
\hspace{2mm}
\includegraphics[width=7.0cm]{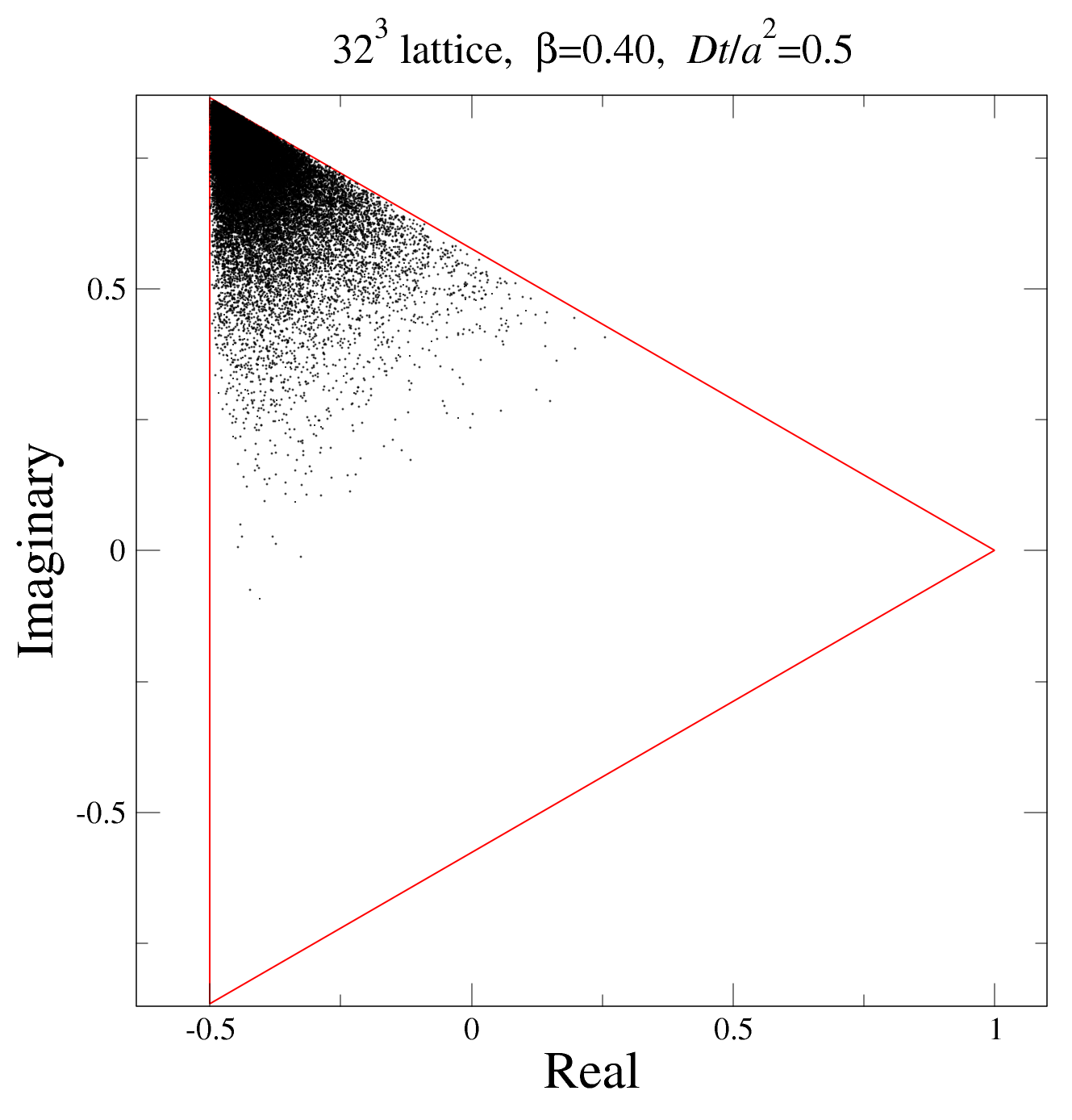}
\vspace{-3mm}
\end{center}
\caption{Distribution of spins $\tilde{s}(\vec{x}, t)$ at each point in one configuration in the broken phase $(\beta=0.40)$ of the three-state Potts model after coarse graining with the diffusion equation.
The diffusion time is $Dt/a^2 =0.1$ (left) and $0.5$ (right). }
\label{fig:potdifhist2}
\end{figure}

Here, we would like to draw a note about spontaneous symmetry breaking.
When symmetry is broken spontaneously in the absence of an external magnetic field, the spin distribution on each configuration becomes asymmetric. 
Figure~\ref{fig:potdifhist2} shows an example where the spin distribution is skewed in the direction of $s = e^{2 \pi i/3}$.
However, due to the $Z_3$ symmetry of this theory, configurations with a skewed distribution toward the three directions of $s = 1, e^{2 \pi i/3}$, and $e^{-2 \pi i/3}$ are always generated with equal probability.
Therefore, if no external field is applied, the expectation value of the magnetization is always zero, even if the symmetry is broken.
The difference between the symmetric and broken cases appears when an infinitesimal external field is applied.
In the symmetric case, the expectation value of the magnetization remains zero, but when the symmetry is broken, even if the external magnetic field is infinitesimal, the expectation value changes in a direction weighted by that external magnetic field.

In order to investigate the correspondence between heavy dense QCD and the Potts model, we replaced $3 \Omega^{\rm (loc)} (\vec{x})$ with spin $s(\vec{s})$.
Let us comment on this correspondence.
Instead of using $\Omega^{\rm (loc)} (\vec{x}) \to s(x)/3$, we can replace it with $\Omega^{\rm (loc)} (\vec{x}) \to b s(x)$, where $b$ is an appropriate real constant.
Then, when $C$ is small, the quark determinant becomes
\begin{eqnarray}
(\mathrm{det}M)^{N_{\rm f}} \rightarrow \prod_{\vec{x}}(1+3bC s (\vec{x}) +3bC^2 s^* (\vec{x}) +C^3 )^{2N_{\rm f}}
\approx 1+ 6N_{\rm f} bC \sum_{\vec{x}} s (\vec{x}). 
\end{eqnarray}
Thus, $h \approx q \approx 6N_{\rm f} bC$, and the slope of the trajectory at $(h,q)=(0,0)$ remains 1 and does not depend on $b$.
When $h$ is large, $Z_3$ symmetry is completely broken, so the nature of the phase transition cannot be discussed from symmetry alone. 
Therefore, the fact that the slope near $(h,q)=(0,0)$ does not depend on $b$ means that $b$ is irrelevant except for quantitative measurements of the critical point.
However, there are limitations to the value of $b$.
If $b=1/2$, then $h$ becomes infinity when $C=1$.
The effective theory with $h= \infty$ does not correspond to QCD, therefore $b$ must be less than $1/2$.
Furthermore, considering the probability distribution of coarse-grained spin, we can see that problems arise when $b \geq 1/2$.
Since the possible values of the coarse-grained spin lie within the triangle in Fig.~\ref{fig:potdifhist1}, if we assign the value $bs(\vec{x},t)$ to $\Omega^{\rm (loc)} (\vec{x})$ when $b \geq 1/2$, then the assigned value will fall outside the range of possible values that $\Omega^{\rm (loc)}$ can take, i.e., the red triangle in Fig.~\ref{fig:pl3hist}.
In such cases, the resulting effective theory will no longer be equivalent to QCD.

\section{Critical points where first-order phase transitions end}
\label{sec:critical}

\subsection{Three-dimensional Potts model with a complex-valued external field}

We study the phase structure of heavy-quark high-density QCD by investigating the three-dimensional three-state Potts model given by Eq.~(\ref{eq:Zpott}).
We fix the hopping parameter at a small value and then gradually increase the chemical potential from a small value to infinity.
The $(h, q)$ parameters we want to investigate are the red curves in Fig.~\ref{fig:hq2} for $N_{\rm f} =2$.
When $C=0$, it is quenched QCD, so the phase transition is first order.
For the case where $h \neq 0$ and $q=0$, this model has been extensively studied, and it is known that the transition changes to a crossover behavior at $h = 0.000517(7)$ \cite{Karsch:2000xv}.
Furthermore, it has been shown that its critical point belongs to the same universality class as the three-dimensional Ising model.
Near $C=0$, $h$ and $q$ are approximately the same, i.e., $h \approx 2N_{\rm f} C$ and $q \approx 2N_{\rm f} C$, so it is expected that the first-order phase transition will change to crossover in the region where $h$ and $q$ are small.

Since the Boltzmann weights of this model are complex numbers, simple Monte Carlo methods cannot be applied.
However, if the value of $q$ is small and the fluctuations in the complex phase are small, the reweighting method can be used.
The reweighting method is a technique that allows us to calculate the expectation value of a physical quantity at a point different from the simulation point, without performing a new simulation, by multiplying a weighting correction factor to the operator before taking the statistical average in Monte Carlo simulations \cite{Ferrenberg:1988yz}.
Let the Boltzmann weight be $e^{-S(\beta, h, q)}$, the original parameters be $(\beta_0, h_0, 0)$, where the Boltzmann weights are real, and the parameters for calculating the expectation value be $(\beta, h, q)$.
Then the expectation value of  $\mathcal{O}$ is given by
\begin{eqnarray}
\langle \mathcal{O} \rangle_{(\beta, h, q)} 
&=& \frac{1}{{\cal Z}(\beta, h, q)} \sum_{s(\vec{x})} \mathcal{O} e^{-S(\beta, h, q)} 
= \frac{\frac{1}{{\cal Z}(\beta_0, h, 0)} \sum_{s(\vec{x})} \mathcal{O} e^{-(S(\beta, h, q)-S(\beta_0, h_0, 0))}e^{-S(\beta_0, h_0, 0)}}{\frac{1}{{\cal Z}(\beta_0, h_0, 0)} \sum_{s(\vec{x})} e^{-(S(\beta, h, q)-S(\beta_0, h_0, 0))}e^{-S(\beta_0, h_0, 0)}} \nonumber \\
&=& \frac{\left\langle \mathcal{O} e^{-(S(\beta, h, q)-S(\beta_0, h_0, 0))} \right\rangle_{(\beta_0, h_0, 0)}}{\left\langle e^{-(S(\beta, h, q)-S(\beta_0, h_0, 0))} \right\rangle_{(\beta_0, h_0, 0)}} .
\end{eqnarray} 
We use the reweighting method to determine the location of the critical point where the first-order phase transition ends.
Moreover, we investigate whether this critical point belongs to the same universality class as the Ising model.

\subsection{Finite volume scaling analysis}
\label{sec:scaling}

We expect that the singular part of the free energy $f= -L^{-d} \ln {\cal Z}$ of the $d$-dimensional Ising model in a space with side length $L$ is expected to satisfy the following scaling law:
\begin{equation}
f(t,h,L^{-1})=L^{-d}f(L^{y_t}t,L^{y_h}h, 1)=L^{-d} \tilde{f}(L^{y_t}t,L^{y_h}h), 
\label{eq:freeeng_u}
\end{equation}
where $t$ represents the reduced temperature, and $h$ represents the external magnetic field.
The critical temperature is $t=0$.
$y_t$ and $y_h$ are the critical exponents with respect to $t$ and $h$, respectively.
The expectation value of magnetization $\langle m \rangle$ is the statistical-mechanical average of the spatial average of the spins, and the magnetic susceptibility $\chi_m$ is
\begin{equation}
  \chi_m(t,0,L^{-1})= L^{d} \langle (m -\langle m \rangle)^2 \rangle .
  \label{eq:susp0}
\end{equation}
From the scaling law of Eq.~(\ref{eq:freeeng_u}), the susceptibility is expressed as the second derivative of the free energy with respect to $h$, 
\begin{equation}
  \chi_m(t,0,L^{-1})= L^{2y_h-d} \tilde{f}_2(L^{y_t}t, 0) .
  \label{eq:susp1}
\end{equation}
Here, $\tilde{f}_2(x,y) = \partial^2 \tilde{f}(x,y)/ \partial y^2$.
Following convention, if we define the critical exponents as $2y_h-d = 2-\eta$ and $y_t = 1/\nu$, 
\begin{equation}
  \chi_m(t,0,L^{-1}) = L^{2-\eta} \tilde{f}_2(L^{1/\nu}t,0) .
  \label{eq:susp2}
\end{equation}
The susceptibility is at its maximum at the critical point, $t=0$, and the magnitude of the peak in susceptibility increases proportionally to $L^{2-\eta}$ and diverges as $L$ approaches infinity.

Moreover, we discuss Binder cumulant $B_4$ to find the critical point \cite{Binder:1981sa}, 
\begin{equation}
B_4 = \frac{\left\langle (m -\langle m \rangle )^4 \right\rangle}{\left\langle (m -\langle m \rangle )^2 \right\rangle^2}
= \frac{ \langle m^4 \rangle_c }{ \langle m^2 \rangle_c } +3 .
\end{equation}
Here,  
$\langle m^2 \rangle_c = \langle (m - \langle m \rangle )^2 \rangle$ and 
$\langle m^4 \rangle_c = \langle (m - \langle m \rangle )^4 \rangle -3 \langle (m - \langle m \rangle )^2 \rangle^2$
are the second-order and fourth-order cumulants, respectively.
Using the scaling law,
\begin{equation}
B_4 (t,0,L^{-1}) = \frac{ L^{4y_h} \tilde{f}_4(L^{y_t}t, 0) }{
 \left( L^{2y_h} \tilde{f}_2(L^{y_t}t, 0) \right)^2} +3
=  \frac{ \tilde{f}_4(L^{y_t}t, 0) }{
 \left( \tilde{f}_2(L^{y_t}t, 0) \right)^2} +3 ,
\end{equation}
where $\tilde{f}_4(x, y) = \partial^4 \tilde{f}(x,y)/ \partial y^4$.
This quantity exhibits no volume dependence ($L$ dependence) at the critical point where $t=0$.
In other words, when we calculate and plot $B_4$ as a function of $t$ for different volumes, we find that for values of $L$ above a certain threshold, the curves of $B_4$ intersect at a single point, which represents the critical point.

\begin{figure}[tb]
\begin{minipage}{0.47\hsize}
\begin{center}
\vspace{0mm}
\includegraphics[width=7.7cm]{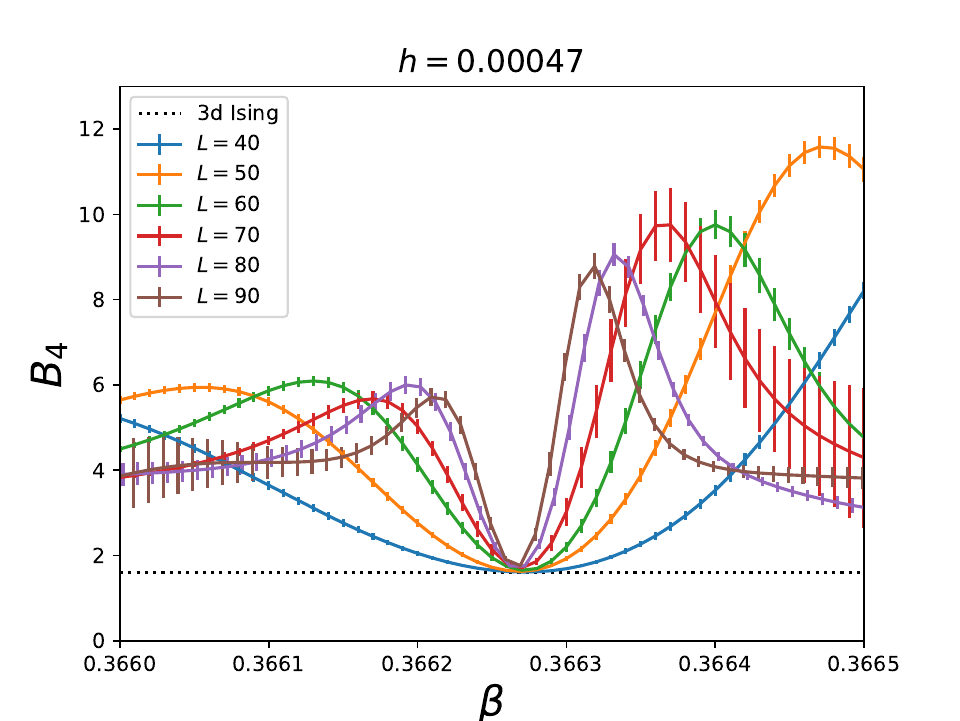}
\vspace{-6mm}
\end{center}
\caption{$\beta$ dependence of the Binder cumulant at $h=0.00047$}
\label{fig:b4beta}
\end{minipage}
\hspace{2mm}
\begin{minipage}{0.47\hsize}
\begin{center}
\vspace{0mm}
\includegraphics[width=7.7cm]{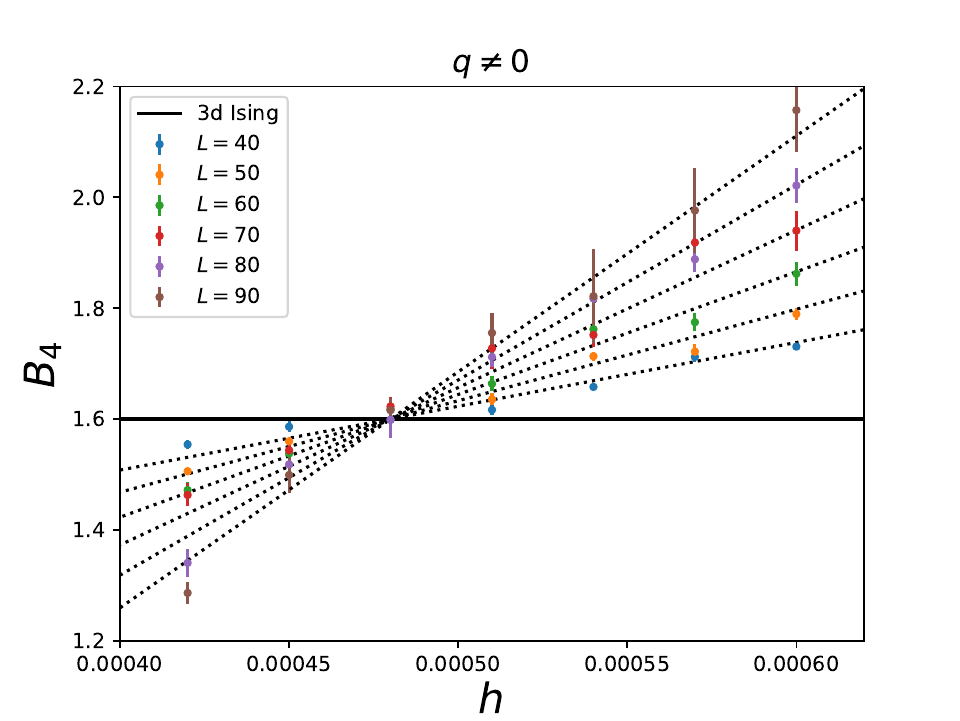}
\vspace{-6mm}
\end{center}
\caption{Binder cumulant at the $\beta_c$ as a function of $h$ on lattices with $L$ from 40 to 90. }
\label{fig:b4scale}
\end{minipage}
\end{figure}

We calculate the magnetization, 
\begin{eqnarray}
m= \frac{1}{N_{\rm site}} \sum_{\vec{x}} \mathrm{Re} [s (\vec{x})] ,
\end{eqnarray}
by Monte Carlo simulation of the three-dimensional three-state Potts model while varying the spatial volume $L^3$.
The standard Metropolis algorithm is used.
For each parameter, the number of independent configurations is set to 200000.
We first determine the $\beta$ at which the phase transition occurs, denoted as $\beta_c$, for various $h$, and then calculate the Binder cumulant at the $\beta_c$ on lattices with side lengths $L$ from 40 to 90.
Figure~\ref{fig:b4beta} shows examples of the $\beta$ dependence of the Binder cumulant near the critical point.
The point where $B_4$ is lowest is $\beta_c$.
In Fig.~\ref{fig:b4scale}, we plot $B_4$ at $\beta_c$ calculated for various $h$ and $L$.
The coefficient $q$ of the imaginary part of the external field term is used to satisfy Eqs. (\ref{eq:pottsh}) and (\ref{eq:pottsq}) with $N_{\rm f}=2$ for each $h$.
It can be seen that $B_4$ as a function of $h$, calculated for all $L$, intersects at a certain $h$.
The point $h$ at which the $L$ dependence disappears is the critical point $h_c$.

In the vicinity of the critical point, by performing a Taylor series expansion of $\tilde{f}(L^{y_t}t,0)$ in $B_4$ with respect to $L^{y_t}t$, we obtain the following equation:
\begin{equation}
B_4 (t,0,L^{-1}) = \frac{ \tilde{f}_4(0, 0) }{
 \left( \tilde{f}_2(0, 0) \right)^2} +3 
+ A t L^{1/\nu} +{\cal O}(t^2).
\end{equation}
$A$ is a proportionality constant.
Assuming $t=h-h_c$, we fit the data as the fitting function 
\begin{equation}
B_4 (h, L) = B_{4c} + A(h-h_c)L^{1/\nu}, 
\end{equation}
where the fit parameters are $B_{4c}$, $h_c$, $\nu$ and $A$.
The fitting results are summarized in Table~\ref{tab:fitpara}.
The dashed lines show the fit functions obtained by fitting the data for each $L$.
The goodness of fit is $\chi^2/{\rm dof} =1.805$.
We also analyzed the case of the standard Potts model where $q$ is set to zero -
the ``$q=0$'' column shows the results for $q=0$, and the ``$q \neq 0$'' column shows the results corresponding to heavy dense QCD.
$B_{4c}$ is the Binder cumulant at the critical point, a quantity that is uniquely determined for each universality class.
Our result, $B_{4c}=1.601(6)$, is consistent with the result of the three-dimensional Ising model, $B_{4c}=1.601$ \cite{Pelissetto:2000ek}.
Moreover, the critical exponent $\nu=0.624(3)$ is slightly different from the value of $\nu=0.630$ for the Ising model, but considering systematic errors, this value can be said to be consistent.
This means that the model we have investigated belongs to the same universality class as the three-dimensional Ising model.

The results for $q=0$, which is the standard three-state Potts model, are consistent with previous studies \cite{Karsch:2000xv}.
Table~\ref{tab:fitpara} shows that the result for the critical point $h_c=0.000479(3)$ of the model corresponding to heavy dense QCD is smaller than that for $q=0$.
Converting to the parameter $C$ of heavy dense QCD, $C$ at the critical point is $C_c = 1.195(7) \times 10^{-4}$.
Since $h$ is small at the critical point, $q$ is also small because $q \approx h$, and the reweighting method works well.

\begin{table}[t]
\begin{center}
\caption{Fitting parameters of the Binder cumulant.}
\label{tab:fitpara}
\begin{tabular}{ccc}
\hline
 &  \hspace{7mm} $q=0$ \hspace{7mm} & $ \hspace{7mm} q\neq0$ \hspace{7mm} \\
\hline
$B_{4c}$ & 1.607(11) &  1.601(6) \\
$h_c$ & 0.000519(5) & 0.000479(3) \\
$\nu$ &  0.628(4) & 0.624(3) \\
$A$ & 2.71(6) & 3.01(1) \\
$\chi^2/{\rm dof}$ & 1.402 & 1.805\\
\hline
\end{tabular}
\end{center}
\end{table}

\begin{figure}[tb]
\begin{minipage}{0.47\hsize}
\begin{center}
\vspace{0mm}
\includegraphics[width=7.7cm]{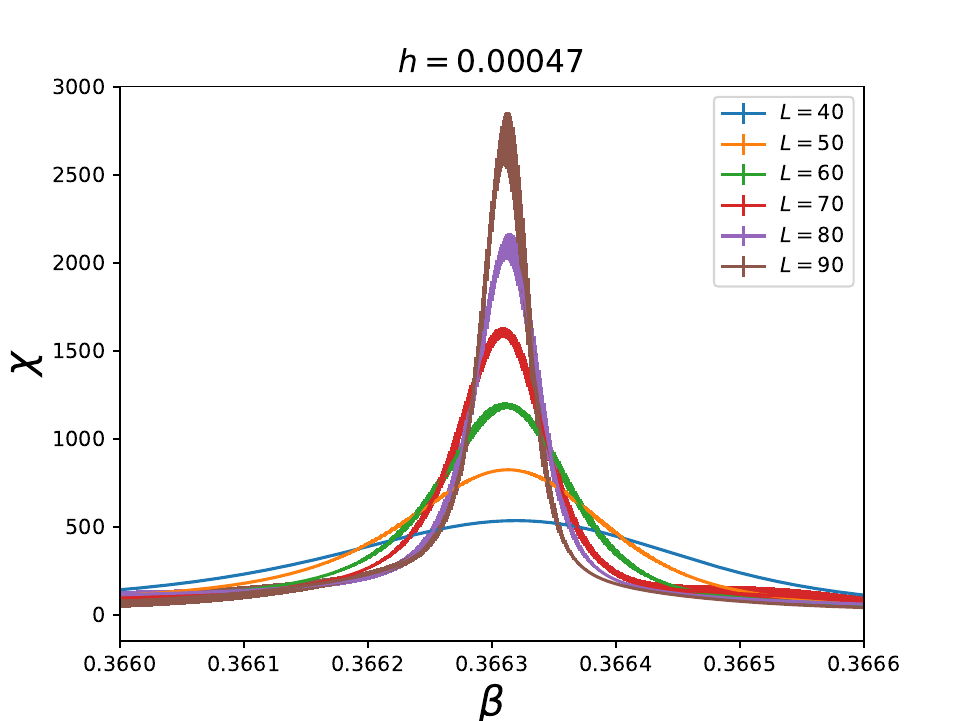}
\vspace{-6mm}
\end{center}
\caption{$\beta$ dependence of the magnetic susceptibility at $h=0.00047$.}
\label{fig:susbeta}
\end{minipage}
\hspace{2mm}
\begin{minipage}{0.47\hsize}
\begin{center}
\vspace{0mm}
\includegraphics[width=7.7cm]{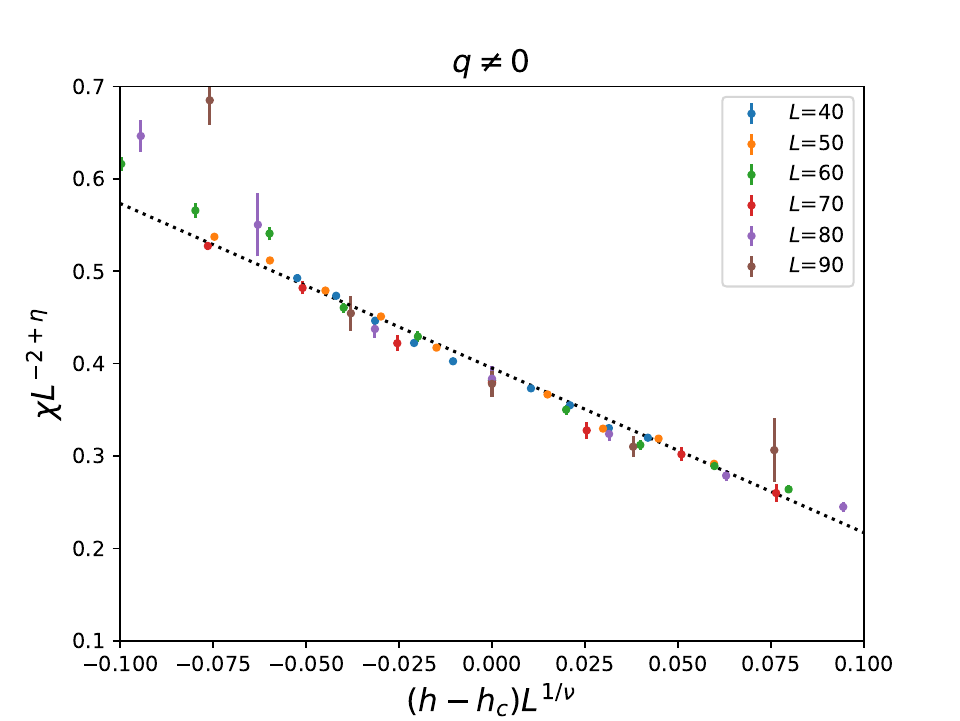}
\vspace{-6mm}
\end{center}
\caption{Scaling plot of the magnetic susceptibility on lattices with $L$ from 40 to 90.}
\label{fig:susscale}
\end{minipage}
\end{figure}

Furthermore, we calculate the magnetic susceptibility $\chi_m$ by Monte Carlo simulation while varying the spatial volume $L^3$.
A reweighting method is used to account for the effect of finite $q$.
Examples of the $\beta$ dependence of susceptibility near the critical point are shown in Fig.~\ref{fig:susbeta}.
The susceptibility peak is located at $\beta_c$, and the height of the peak increases as $L$ increases.
By performing a Taylor expansion of the derivative of the free energy Eq.(\ref{eq:susp2}) around the critical point $t=0$, with respect to $L^{1/\nu} t$, we obtain the following scaling law:
\begin{equation}
  \chi_m(t,0,L^{-1}) L^{-2+\eta} =  \tilde{f}_2(0,0) + B t L^{1/\nu} +{\cal O}(t^2), 
  \label{eq:susp2fit}
\end{equation}
where $B$ is a proportionality constant.
Assuming $t=h-h_c$, the scaling plot, $(h-h_c) L^{1/\nu}$ vs. $\chi_m L^{-2+\eta}$, is shown in Fig.~\ref{fig:susscale}.
In this plot, the critical exponents $\nu$ and $\eta$ are those of the Ising model, and the critical point $h_c$ is the result obtained from the fit of $B_4$.
From Fig.~\ref{fig:susscale}, we can see that the data are aligned in a straight line near the critical point.
The dashed line is a linear fit to the data near the critical point.
This scaling plot also shows that the phase transition of this model, which corresponds to heavy dense QCD, belongs to the same universality class as the three-dimensional Ising model.

\section{Crossover region at high density}
\label{sec:crossover}

\subsection{Avoiding the sign problem}

\begin{figure}[tb]
\begin{center}
\vspace{0mm}
\includegraphics[width=8.0cm]{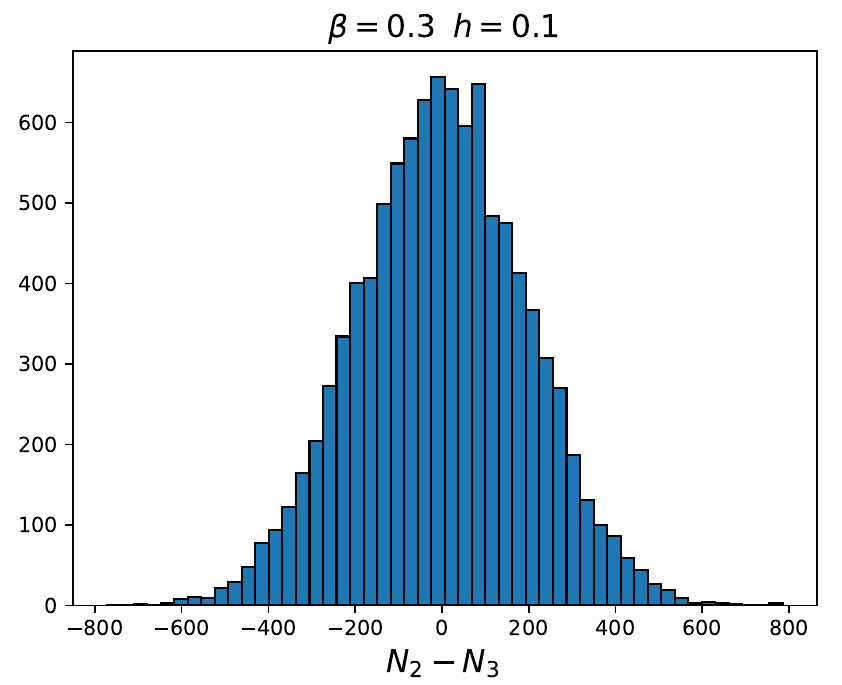}
\vspace{-3mm}
\end{center}
\caption{Probability distribution of $N_2-N_3$ at $\beta =0.3$ on a $32^3$ lattice.
}
\label{fig:n2n3dis}
\end{figure}

We compute physical quantities in the crossover region by the reweighting method.
When the coefficient $q$ of the imaginary part of the external field term is large and the volume is large, the sign problem is severe.
The imaginary terms are expressed as the number of spins:
\begin{eqnarray}
q\sum_{\vec{x}} \mathrm{Im} [s (\vec{x})]
=q \frac{\sqrt{3}}{2} (N_2 - N_3) \equiv q \phi. 
\end{eqnarray}
Using the reweighting method, the magnetization $\langle m \rangle_q$ and the energy $\langle E \rangle_q$, incorporating the complex external field, are
\begin{eqnarray}
\langle m \rangle_q = \frac{\langle m \cos (q \phi) \rangle_0}{
\langle \cos (q \phi) \rangle_0} \ \ {\rm and} \ \
\langle E \rangle_q = \frac{\langle E \cos (q \phi) \rangle_0}{
\langle \cos (q \phi) \rangle_0}.
\label{eq:mrewighting}
\end{eqnarray}
Here, the magnetization is
\begin{eqnarray}
m= \frac{1}{N_{\rm site}} \left( N_1 - \frac{1}{2}(N_2+N_3) \right) 
= \left( \frac{3 N_1}{2N_{\rm site}} -\frac{1}{2} \right),
\label{eq:m-n1}
\end{eqnarray}
and the energy operator is defined by
\begin{eqnarray}
E= - \sum_{\vec{x}} \sum_{i=1}^3 \mathrm{Re} \left[ s (\vec{x}) s^* (\vec{x}+\hat{i}) \right].
\end{eqnarray}

Introducing probability distribution functions for $N_1$, $N_2$, and $N_3$, $W(N_1, N_2, N_3)$, the magnetization can be rewritten as 
\begin{eqnarray}
\langle m \rangle_q = \frac{
\sum_{N_1, N_2, N_3} m \cos (q \phi) \, W(N_1, N_2, N_3) }{
\sum_{N_1, N_2, N_3} \cos (q \phi) \, W(N_1, N_2, N_3) } .
\label{eq:mcosw}
\end{eqnarray}
As seen in Eq.~(\ref{eq:m-n1}), $m$ depends only on $N_1$.
Also, since $\phi$ depends only on $N_2 -N_3$, we first sum over $N_1$ in Eq.~(\ref{eq:mcosw}).
Then, if we assume that $N_2$ and $N_3$ are distributed randomly, the distribution functions for $N_2$ and $N_3$ can be assumed to be Gaussian, i.e.,
\begin{eqnarray}
\sum_{N_1} W(N_1, N_2, N_3) \propto e^{-\frac{3}{16\alpha_1} (N_2 -N_3)},
 \hspace{5mm}
\sum_{N_1} m W(N_1, N_2, N_3) \propto e^{- \frac{3}{16\alpha_2} (N_2 -N_3)},
\end{eqnarray}
where $\alpha_1$ and $\alpha_2$ are appropriate constants.
A typical example of the distribution of $\sum_{N_1} W(N_1, N_2, N_3)$ as a function of $N_2-N_3$ is shown in Fig.~\ref{fig:n2n3dis}.
The distribution appears to be Gaussian \cite{Ejiri:2007ga,Ejiri:2009hq}.
If $N_{\rm site}$ is large enough that $\phi/N_{\rm site} = (\sqrt{3}/2) (N_2-N_3)/N_{\rm site}$ can be considered a continuous variable, we get the following equation:
\begin{eqnarray}
\langle m \rangle_q \propto \frac{
\sum_{N_2, N_3} \cos \left[ q \frac{\sqrt{3}}{2} (N_2 - N_3) \right] e^{-\frac{3}{16\alpha_2} (N_2 -N_3)} }{
\sum_{N_2, N_3} \cos \left[ q \frac{\sqrt{3}}{2} (N_2 - N_3) \right] e^{- \frac{3}{16\alpha_1} (N_2 -N_3)} }
\propto e^{(\alpha_1-\alpha_2) q^2} .
\end{eqnarray}
Therefore, if we write $\langle m \rangle_q \equiv \langle m \rangle_0 e^{f_m (q)}$, $f_m (q)$ is a function proportional to $q^2$.
A similar argument can be made for $\langle E \rangle_q$ by introducing the probability distribution function $W(N_1,N_2,N_3, E)$ for $N_1, N_2, N_3,$ and $E$.
We denote  $\langle E \rangle_q \equiv \langle E \rangle_0 e^{f_E (q)}$. 
If we assume that the complex phase is Gaussian distributed,
$f_E (q)$ is a function proportional to $q^2$.

\begin{figure}[tb]
\begin{center}
\vspace{0mm}
\includegraphics[width=7.7cm]{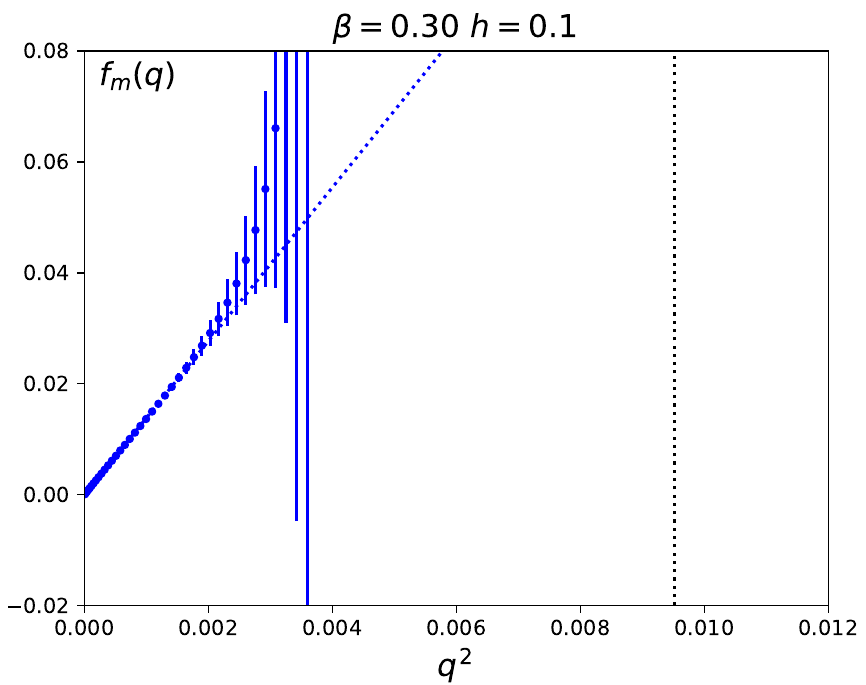}
\hspace{2mm}
\includegraphics[width=7.7cm]{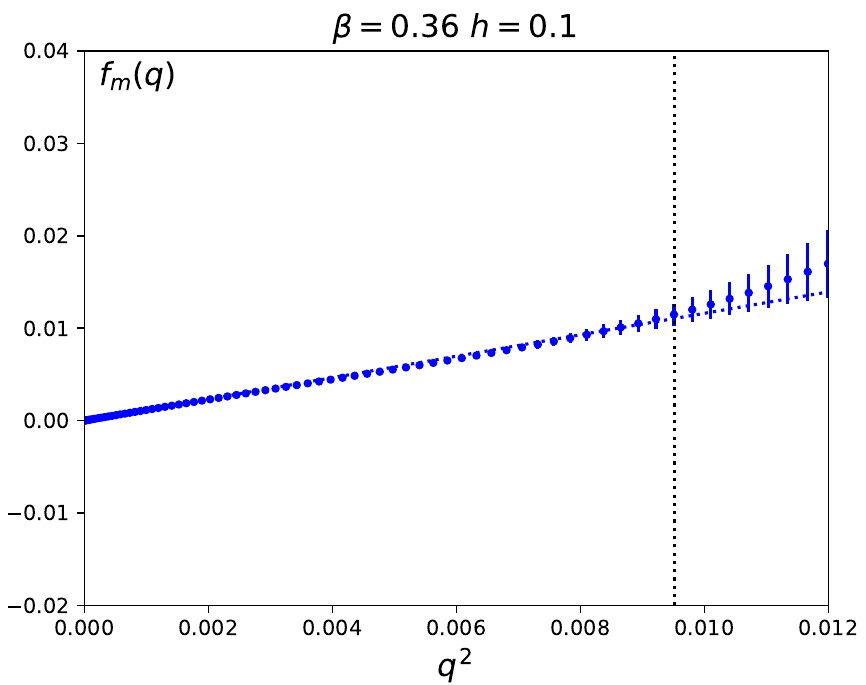}
\vspace{-3mm}
\end{center}
\caption{$q$ dependence of $f_m (q)$. 
The left panel shows the results for $(\beta, h) = (0.30, 0.1)$, where the sign problem is severe, and the right panel shows the results for $(\beta, h) = (0.36, 0.1)$, where the sign problem does not occur.
}
\label{fig:fmq2}
\end{figure}

\begin{figure}[tb]
\begin{center}
\vspace{0mm}
\includegraphics[width=7.7cm]{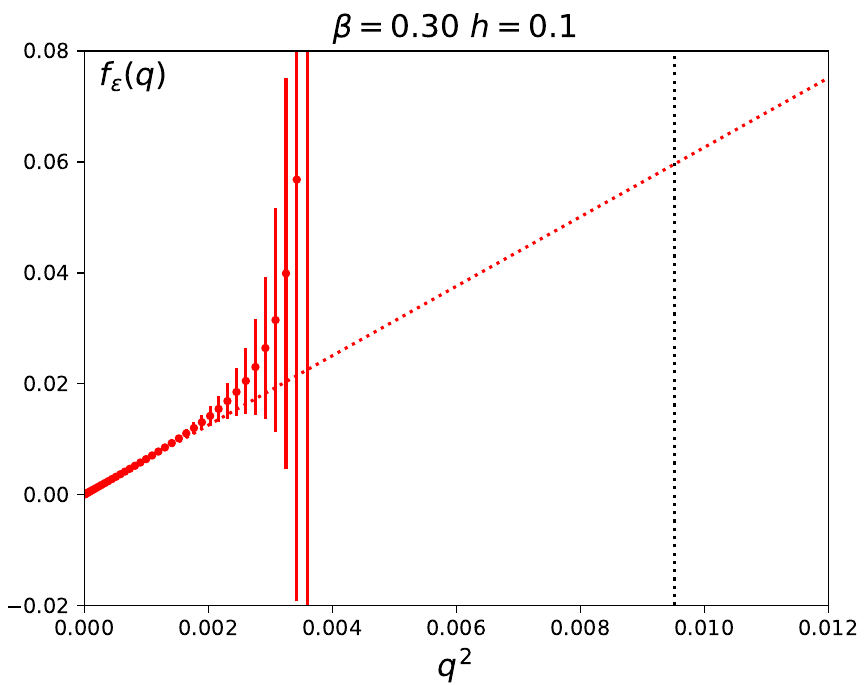}
\hspace{2mm}
\includegraphics[width=7.7cm]{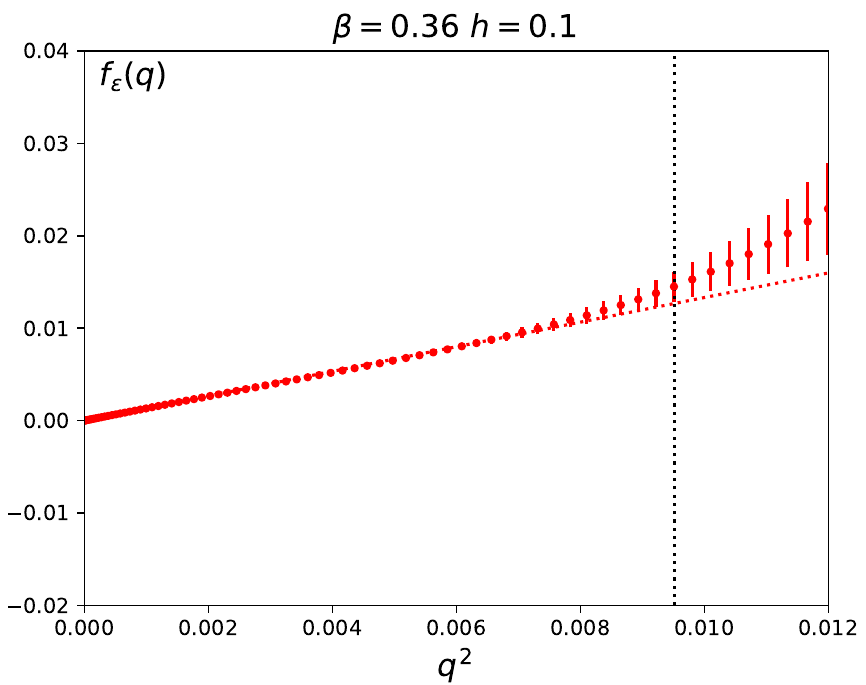}
\vspace{-3mm}
\end{center}
\caption{$q$ dependence of $f_E (q)$. The left and right panels show the results for $(\beta, h) =(0.30, 0.1)$ and  $(0.36, 0.1)$, respectively.
}
\label{fig:feq2}
\end{figure}

We plot examples of the $q$ dependence of $f_m (q)$ and $f_E (q)$ in Fig.~\ref{fig:fmq2} and Fig.~\ref{fig:feq2}, respectively.
The simulations are performed on a $16^3$ lattice.
The left figure shows the results when $\beta$ is small, which causes serious sign problems, while the right figure shows the results when $\beta$ is large, which causes no sign problems.
The vertical dashed lines shown in Figs.~\ref{fig:fmq2} and \ref{fig:feq2} represent the parameter $q$ corresponding to heavy-quark QCD.
The horizontal axis is $q^2$ and both figures are consistent with being proportional to $q^2$ within statistical error.
In the region of $q$ where the statistical error is large, even a slight change in $\beta$ causes these center values to bend above or below the straight line, so even assuming a fourth-order or higher function, this calculation was unable to determine the coefficient of $q^4$.

If the sign problem is serious, the values of $f_m (q)$ and $f_E (q)$ at the desired $q$ cannot be calculated because the statistical error becomes too large.
Therefore, to calculate the magnetization and energy, we fit the data at small $q$ assuming that they are proportional to $q^2$, and then extrapolate the values of $f_m (q)$ and $f_E (q)$ up to the desired $q$.

\subsection{Energy and magnetization at large $h$ and $q$}

We use the reweighting method to calculate the expectation values of the energy and magnetization in the presence of a complex external field whose partition function is given by Eq.~(\ref{eq:mrewighting}).
Since we found that the spatial volume dependence is small in the three-dimensional three-state Potts model except near the critical point, we perform Monte Carlo simulations using a relatively small lattice size $N_{\rm site}=16^3$.
The sign problem in the reweighting method becomes more serious as the volume increases, but it is not so serious at that lattice size.
The number of independent configurations is 1000000 for each parameter.

The expectation value of the magnetization, which is the order parameter, is shown in Fig.~\ref{fig:magneb}.
The horizontal axis is $\beta$ and $h$ is set to $0.0, 0.1, 0.5,$ and $1.0$.
The results when the complex external field coefficients $h$ and $q$ are given by Eqs.~(\ref{eq:pottsh}) and (\ref{eq:pottsq}) with $N_{\rm f}=2$ are shown by crosses, and for comparison, the cases where $q=0$ are plotted by circles.
First, it can be seen that the effect of the imaginary part of the external field, i.e., the term proportional to $q$, is small.
In other words, using the extrapolation approximation of $q^2$ in the reweighting method does not significantly affect the quantitative results.
When $h$ becomes larger than the critical point determined in Sec.~\ref{sec:critical} and $h$ is further increased, the change becomes milder.
Figure~\ref{fig:magneb} shows that  the phase transition point $\beta_c$ decreases as $h$ increases, and eventually the phase transition itself disappears.
Similarly, we plot the expectation value of the energy density, $\varepsilon = E/ N_{\rm site}$, in Fig.~\ref{fig:energb}.
As with the magnetization, the change in the energy density becomes slower as $h$ increases, and eventually the phase transition disappears.

\begin{figure}[tb]
\begin{minipage}{0.47\hsize}
\begin{center}
\vspace{0mm}
\includegraphics[width=7.7cm]{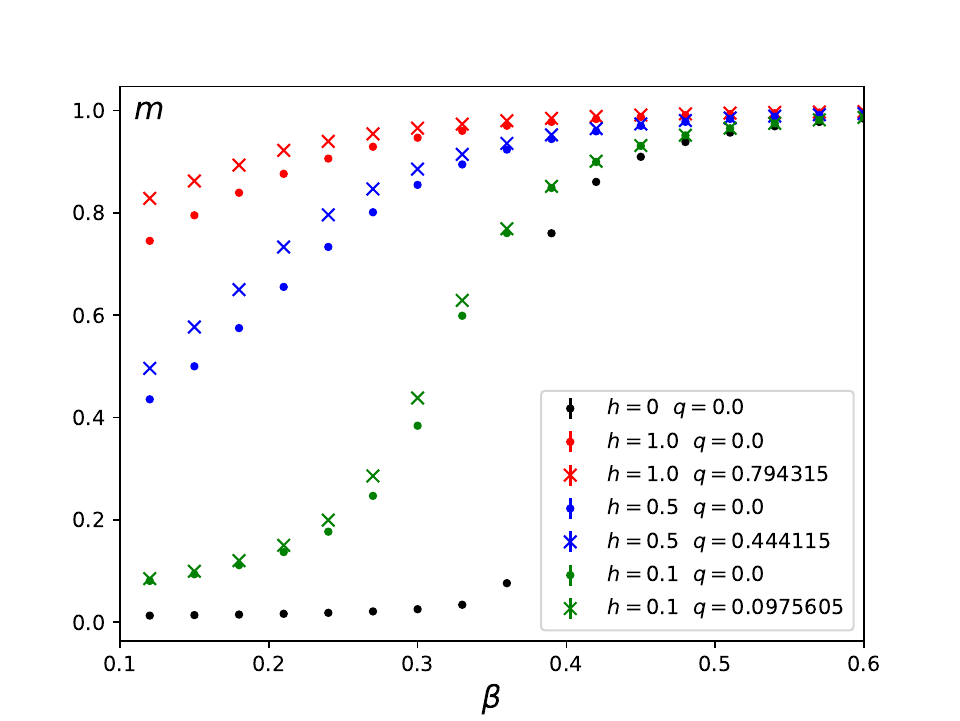}
\vspace{-6mm}
\end{center}
\caption{Magnetization as a function of $\beta$ and $h$ on a $16^3$ lattice.}
\label{fig:magneb}
\end{minipage}
\hspace{2mm}
\begin{minipage}{0.47\hsize}
\begin{center}
\vspace{0mm}
\includegraphics[width=7.7cm]{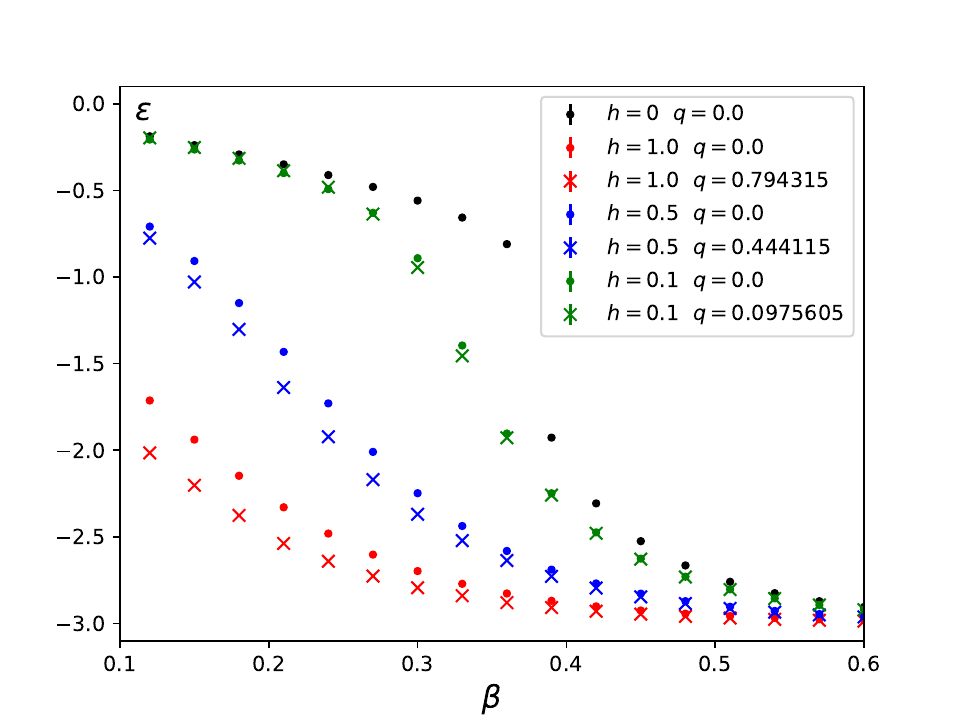}
\vspace{-6mm}
\end{center}
\caption{Energy density as a function of $\beta$ and $h$ on a $16^3$ lattice.}
\label{fig:energb}
\end{minipage}
\end{figure}

\subsection{Comparison with heavy dense QCD}

When the hopping parameter $\kappa$ in QCD is small, $C$ is a monotonically increasing function of $\mu$,
i.e., $C=(2\kappa)^{N_t} e^{\mu/T}$.
When $C$ is increased from $C=0$, the phase transition is first order up to the critical point, after which it becomes a crossover.
Since $h$ increases up to $C=1$, as $C$ increases further the change in the order parameter becomes milder and eventually the phase transition disappears.
After $C=1$, $h$ becomes smaller and the phase transition becomes stronger as $C$ approaches the second critical point.
Once the critical point is passed, the phase transition is first order up to infinity.
In Sec.~\ref{sec:scaling}, we found the critical point is $C_c = 1.195(7) \times 10^{-4}$.
At the same time, $C_c=1/[1.195(7)] \times 10^{4}$ is also the critical point.
Since the critical chemical potential is given by $\mu_c/T = \ln C_c - N_t \ln(2 \kappa)$, increasing $\kappa$ decreases $\mu_c$.
When the quark mass is reduced (when $\kappa$ is increased), the smaller $\mu_c$ quickly becomes zero, but it is interesting to see how the larger $\mu_c$ changes.

The critical point, where $C$ is very large, is related to the filling of space with quarks.
When $C$ is infinite, the number of quarks reaches a maximum and fills space.
In heavy-quark high-density QCD, when $C$ is infinite, the quark determinant becomes a constant, which is the same as in quenched QCD.
Therefore, heavy dense effective theory has symmetry under the transformation from $C$ to $1/C$.
However, the critical point in the large $C$ region is not thought to be directly connected to the critical point in finite-density QCD for physical quark masses, which is of experimental interest.
This is because the critical point at that physical point is not expected to be at a density where quarks fill space.
Furthermore, the situation where space is filled with quarks is related to the $N_{\rm site}$.
If the physical volume is fixed and the lattice spacing is changed, the $N_{\rm site}$ changes, so the critical point in the large $C$ region depends on the lattice spacing.

If $C$ is small, the quark determinant in heavy dense QCD is
\begin{eqnarray}
\mathrm{det}M \approx \prod_{\vec{x}} \left[1
+3C \Omega^{\rm (loc)} (\vec{x}) \right]^2
\approx \prod_{\vec{x}} \exp \left[ 6C \Omega^{\rm (loc)} (\vec{x}) \right] 
= \exp (6C N_s^3 \Omega), 
\end{eqnarray}
where $\Omega$ is the Polyakov loop.
An approximation that ignores the complex phase part of $\det M$ allows Monte Carlo simulation.
In Ref.~\cite{Ejiri:2025fsf}, we performed Monte Carlo simulations for the lattice action,
\begin{eqnarray}
S=S_g + {\rm Re} \ln \det M = 6\beta N_s^3 N_t P + 6C N_s^3 {\rm Re} \Omega, 
\end{eqnarray}
where $P$ is the plaquette.
We found that as $C$ increases, the $\beta$ at which the phase transition occurs decreases, and the change in the expectation value of the plaquette becomes sharper.
The reason for this sharp change is the behavior in the strong coupling region where $\beta$ is small.
In the case of this lattice action, in the strong coupling region, the plaquette behaves as follows up to $O(\beta^2)$:
\begin{eqnarray}
\langle P \rangle = \frac{\beta}{18} + \frac{\beta^2}{216}.
\end{eqnarray}
Over a relatively wide range where $\beta$ is small, the behavior of the plaquette expectation value is consistent with this equation \cite{Ejiri:2025fsf}.
The effect of the $\Omega$ term in the action only appears in terms $O(\beta^{N_t-1})$ and above.
The strong coupling region is the confinement phase, and as $C$ increases, $\beta$ at the phase transition point decreases. 
Therefore, when $\langle P \rangle$ is small, the phase transition occurs and $\langle P \rangle$ becomes a large value in the deconfinement phase, so the change in $\langle P \rangle$ becomes steeper as $C$ increases.
Although we must add the effect of the complex phase of $\det M$, this behavior suggests the existence of a new singularity.

On the other hand, in the three-state Potts model, no new singularities appear with increasing $C$ or $h$.
In the Potts model, the quantity corresponding to the plaquette in QCD is the expectation value of $-1$ times the energy, but in the small $\beta$ region, the energy behaves quite differently from the plaquette.
The result of the strong coupling expansion up to the first order in $\beta$ and the second order in $h, q$ is the following equation:
\begin{eqnarray}
\frac{\langle E \rangle}{N_{\rm site}} =
\frac{3}{4}q^2 -\frac{3}{4}h^2 -\frac{9}{8}h q^2 - \frac{9}{32}h^2q^2 +
\beta \left( -\frac{3}{2} + \frac{33}{8} q^2 - \frac{33}{8} h^2 - \frac{189}{16} h q^2 -\frac{183}{64} h^2q^2 \right).
\hspace{6mm}
\end{eqnarray}
See Appendix~\ref{sec:hightemp} for details.
The zeroth-order term of $\beta$ in $\langle E \rangle$ varies depending on $h$ and $q$ to approach the value of the deconfined phase, and $h$ acts to weaken the phase transition even in the region where $\beta$ is small.
In the process of simplifying from the heavy dense QCD to the three-state Potts model as the effective theory, the property of the rapid change of the plaquette seems to have been lost.

\section{Phase structure with complex chemical potential} 
\label{sec:complexmu}

Finally, we consider the case where the chemical potential is complex, 
$\mu= \mu_{\rm R} +i \mu_{\rm I}$.
If we define real parameters as $C=(2\kappa)^{N_t} e^{\mu_{\rm R}/T}$, 
the quark matrix of the heavy dense QCD becomes
\begin{eqnarray}
\det M = \prod_{\vec{x}} \left\{1
+ 3C e^{i \mu_{\rm I}/T} \Omega^{\rm (loc)} (\vec{x}) 
+ 3C^2 e^{2i \mu_{\rm I}/T} \left( \Omega^{\rm (loc)} (\vec{x}) \right)^* 
+ C^3 e^{3i \mu_{\rm I}/T} \right\}^2.
\end{eqnarray}
As in Sec.~\ref{sec:eff3dmodel}, we can construct the corresponding three-state spin model, but no simple formula is obtained.
We, therefore, discuss the important but simple cases $\mu_{\rm I}/T =2 \pi /3$ and $\mu_{\rm I}/T = \pi /3$.
Although the real $\mu$ case is not much different from $\mu=0$, interesting singularities appear for complex $\mu$.

QCD has the $Z_3$ center symmetry under the transformation from $\Omega^{\rm (loc)} (\vec{x})$ to 
$ e^{2 \pi i/3 } \Omega^{\rm (loc)} (\vec{x})$, except for the quark determinant.
The quark determinant for $\mu_{\rm I}/T =2 \pi /3$ is obtained by replacing $\Omega^{\rm (loc)} (\vec{x})$ in $\det M$ for $\mu_{\rm Im} =0$ with $e^{2 \pi i/3 } \Omega^{\rm (loc)} (\vec{x})$.
Here, if we consider $e^{2 \pi i/3 } \Omega^{\rm (loc)} (\vec{x})$ as a new $\Omega^{\rm (loc)} (\vec{x})$, the theory itself does not change \cite{Roberge:1986mm}.
When considering the corresponding spin model, if we consider $e^{2 \pi i/3 } \Omega^{\rm (loc)} (\vec{x})$ as $s(x)$, the effective model is the same.

When $\mu_{\rm I}/T = \pi /3$, the quark determinant is
\begin{eqnarray}
\det M &=& \prod_{\vec{x}} \left\{1
+ 3C e^{\pi i/3} \Omega^{\rm (loc)} (\vec{x}) 
+ 3C^2 e^{2\pi i/3} \left( \Omega^{\rm (loc)} (\vec{x}) \right)^* 
+ C^3 e^{3\pi i/3} \right\}^2 \nonumber \\
&=& \prod_{\vec{x}} \left\{1
- 3C e^{-2\pi i/3} \Omega^{\rm (loc)} (\vec{x}) 
+ 3C^2 \left( e^{-2\pi i/3}\Omega^{\rm (loc)} (\vec{x}) \right)^* 
- C^3 \right\}^2
\end{eqnarray}
Using $Z_3$ symmetry, we consider the phase-transformed Polyakov loop, $e^{-2\pi i/3} \Omega^{\rm (loc)} (\vec{x})$, as a spin $s(x)$, and carry out the same discussion as in Sec.~\ref{sec:eff3dmodel}, 
the spin model corresponding to $\mu_{\rm I}/T = \pi /3$ is obtained by replacing $C$ in Eqs.~(\ref{eq:Zpott}), (\ref{eq:pottsh}), and (\ref{eq:pottsq}) with $-C$.
When $C$ is small near $(h, q)=(0,0)$, $h$ and $q$ behave as 
$h \approx -2 N_{\rm f} C$ and $q \approx -2 N_{\rm f} C$.
$h$ changes in the negative direction.\footnote{
The three-state Potts model in the presence of a negative external field is investigated in Ref.~\cite{Bonati:2010ce}.}

\begin{figure}[tb]
\begin{minipage}{0.47\hsize}
\begin{center}
\vspace{0mm}
\includegraphics[width=7.7cm]{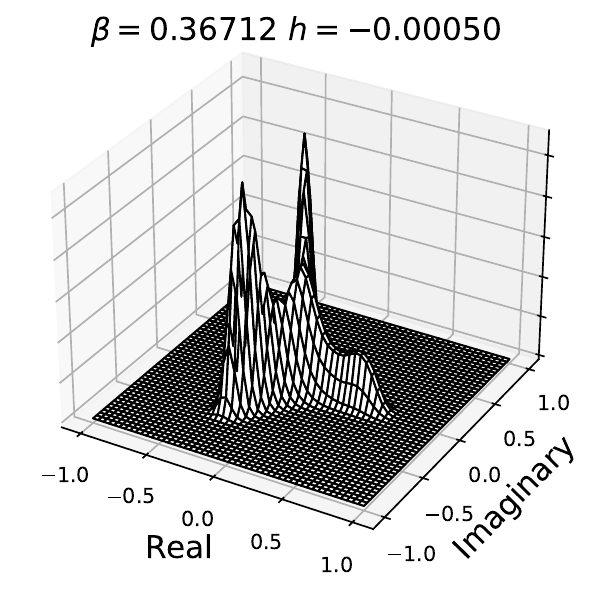}
\vspace{-9mm}
\end{center}
\caption{
Probability distribution of the magnetization of the three-state Potts model with $h=-0.0005$ and $q=0$.}
\label{fig:plhishneg}
\end{minipage}
\hspace{2mm}
\begin{minipage}{0.47\hsize}
\begin{center}
\vspace{0mm}
\includegraphics[width=7.7cm]{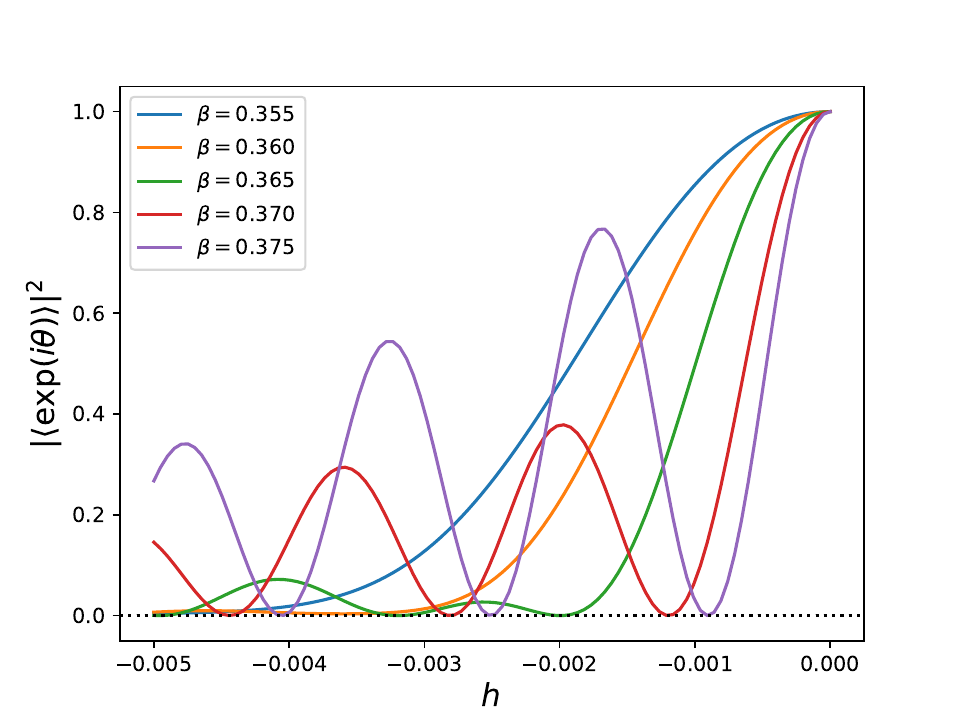}
\vspace{-6mm}
\end{center}
\caption{$| {\cal Z}(\beta, h, q) /{\cal Z}(\beta, h, 0) |^2 $ as a function of $h$ for various $\beta$.}
\label{fig:zzero}
\end{minipage}
\end{figure}

We show the histogram of the magnetization in the complex plane measured at  $(\beta, h) =(0.36712, -0.0005)$ on a $16^3$ lattice in Fig.~\ref{fig:plhishneg}, which changes from the histogram for $(h, q) = (0, 0)$ to that for $h$ in the negative direction by the reweighting method with $q=0$.
The two peaks at ${\rm Re} (m) <0$ become larger. 
Since the heights of the peaks are the same, the partition function is the sum of the contributions of the two peaks.
Because $q$ is nonzero, the Boltzmann weights have complex phases, and the partition function ${\cal Z}$ is zero at the point where the difference in complex phase between the two peaks is exactly an odd multiple of $\pi$.
This is called the Lee-Yang singularity \cite{Yang:1952be,Lee:1952ig}.
At the point where ${\cal Z} = 0$, the free energy diverges.

We calculate the ratio of the partition functions at $q \neq 0$ and $q=0$ by simulations \cite{Ejiri:2005ts}
\begin{eqnarray}
\left| \frac{{\cal Z}(\beta, h, q)}{{\cal Z}(\beta, h, 0)} \right|
= \left| \langle \exp (i\theta) \rangle_{(\beta, h, 0)} \right|
= \sqrt{ \langle \cos \theta \rangle_{(\beta, h, 0)}^2 + \langle \sin \theta \rangle_{(\beta, h, 0)}^2 }
\end{eqnarray}
with $\theta = q\sum_{\vec{x}} \mathrm{Im} [s (\vec{x})] $.
$\langle \cdots \rangle_{(\beta, h, q)}$ means the expectation value measured at $(\beta, h, q)$.
Because ${\cal Z}(\beta, h, 0) > 0$, if the ratio is zero, then ${\cal Z}(\beta, h, q) =0$.
We use $h$ and $q$ when $C$ in Eqs.~(\ref{eq:pottsh}) and (\ref{eq:pottsq}) is replaced by $-C$.
We plot $| {\cal Z}(\beta, h, q) /{\cal Z}(\beta, h, 0) |^2$ measured on a $16^3$ lattice in Fig.~\ref{fig:zzero} for various $\beta$.
When $\beta \ge 0.365$, singularities where ${\cal Z}=0$ appear periodically.

The fact that a singularity appears only when $\beta \ge 0.365$ can be understood as follows.
When $h=0$, in the symmetric phase with small $\beta$, the distribution peak is at the origin, while in the broken phase with large $\beta$, three peaks with $Z_3$ symmetry appear.
At the first-order phase transition point, the distribution function has four peaks: one for the symmetric phase and three for the broken phase.
When $h$ is decreased from $h=0$, the probability distribution for the imaginary axis direction does not change, but the probability along the real axis increases in the negative direction because of the weight $e^{h\sum_{\vec{x}} \mathrm{Re} [s (\vec{x})]}$.
As a result, in the broken phase $(\beta=0.370, 0.375)$ and near the first-order phase transition $(\beta=0.365)$, two peaks become larger and of the same height, as shown in Fig.~\ref{fig:plhishneg}.
As $|q|$ increases, the complex phase difference between the two peaks becomes larger.
At the point $(h,q)$ where the difference between these complex phases $\theta$ is an odd multiple of $\pi$, ${\cal Z}$ becomes zero.
As $\beta$ increases, the distance between the two peaks increases, and, therefore, the interval of $q$ (or $h$) where  ${\cal Z} =0$ becomes narrower, as shown in Fig.~\ref{fig:plhishneg}.
Furthermore, since $\theta$ is proportional to the volume, increasing the spatial volume narrows the interval of $q$ at which ${\cal Z} =0$.
This singularity corresponds to the Roberge-Weiss singularity in finite-density lattice QCD \cite{Roberge:1986mm}.
On the other hand, in the symmetric phase $(\beta=0.355, 0.360)$, there is only one peak in the histogram, so ${\cal Z}$ does not become 0 and decreases exponentially as $(h, q)$ increases.

\section{Conclusions and outlook}
\label{sec:summary}

We discussed the phase structure of dense QCD with heavy dynamical quarks through the effective theory of the three-state Potts model.
When we perform a hopping parameter expansion for the quark determinant, the properties of the expansion coefficients allow us to include the effect of the spatial link field in the Polyakov loop term, which is made up of only the link field in the time direction, by shifting $\kappa$ to $\kappa_{\rm eff}$.
We then discuss an effective theory for investigating high-density regions, ignoring the spatial link field.
Furthermore, we corresponded high-density QCD to the three-dimensional three-state Potts model with the same symmetry breaking.

The Potts model has a complex-valued external field, and we investigated the phase transition with parameters corresponding to heavy dense QCD.
At low densities, the transition is first order, but as the density increases, the transition changes to a crossover at a critical point, weakening the transition.
As the density increases further, a critical point appears, and above that point, the transition becomes first order.
The critical point belongs to the three-dimensional Ising universality class.

In the heavy-quark high-density limit, the quark determinant becomes a constant, the same as in quenched QCD.
Therefore, the high-density limit of QCD becomes a first-order phase transition.
In the high-density limit, quarks fill up space.
The second critical point at high densities is thought to be related to this filling of space with quarks.
If this is the case, it is likely unrelated to the first-order phase transition at finite density that is of interest in experiments.
We were interested in whether singularities would appear in the region where $h$ is large.
However, we found that singularities do not appear when the external field is large in the three-state Potts model with a complex external field examined in this study.

The three-state Potts model, which corresponds to high-density QCD and which we have introduced here, is a three-dimensional theory, so it is easy to analyze, and in particular, it can be analyzed using the tensor renormalization group, which does not have a sign problem.
The Potts model has the same broken symmetry as QCD in the heavy-quark region, and so the behavior of the order parameters is also very similar.
Studying this effective model may contribute to QCD.

\section*{Acknowledgments} 
We thank Toshiki Sato for fruitful discussions.
This work is supported by JSPS KAKENHI (Grants No. JP21K03550, and No. JP25K07299).

\appendix

\section{High-temperature expansion of the three-state Potts model} 
\label{sec:hightemp}

At high temperatures, assuming $\beta$ is small, we expand ${\cal Z}$ up to first order in $\beta$, 
\begin{eqnarray}
{\cal Z} &=& \sum_{s(\vec{x})} \left[
1+\beta\sum_{\vec{x}} \sum_{i=1}^3 \mathrm{Re}(s(\vec{x}) s^*(\vec{x}+\hat{i}) 
+O(\beta^2) \right]
\exp \left[
h\sum_{\vec{x}} \mathrm{Re}(s(\vec{x})] + iq \sum_{\vec{x}} \mathrm{Im}(s(\vec{x})) 
\right] \nonumber \\
&=& f^{N_{\rm site}} + 3 \beta N_{\rm site} f^{N_{\rm site} -2} g +O(\beta^2), 
\end{eqnarray}
where
\begin{eqnarray}
f &=& e^h +2 e^{-h/2} \cos \left( \frac{\sqrt{3}}{2} q \right) , \\
g &=& e^h -e^{-h} -2 e^{h/2} \cos \left( \frac{\sqrt{3}}{2} q \right) 
+ 2e^{-h} \cos \left( \sqrt{3} q \right) .
\end{eqnarray}
$\sum_{s(\vec{x})}$ means the sum of the values of 
$s(\vec{x}) =\{1, e^{2 \pi i/3}, e^{4 \pi i/3} \}$ at each point $\vec{x}$.
The magnetization is calculated by differentiating the partition function with respect to $h$,
\begin{eqnarray}
m &=& \frac{1}{N_{\rm site}} \frac{\partial}{\partial h}{\cal Z} \nonumber \\
&=& \frac{1}{1+3\beta N_{\rm site} f^{-2} g} \left[
\left( f^{-1} + 3\beta (N_{\rm site} -2) f^{-3} g \right) \left( e^h - e^{-h/2} \cos \left(\frac{\sqrt{3}}{2} q\right) \right) 
\right. \nonumber \\ 
&& \left. +3 \beta f^{-2} \left( 2e^{2h} +e^{-h} -e^{h/2} \cos \left( \frac{\sqrt{3}}{2} q \right) 
-2e^{-h} \cos \left(\sqrt{3} q \right) \right)
\right]
\end{eqnarray}
Expanding to the first order in $\beta$ and the second order in $h,q$,
\begin{eqnarray}
m= 
\beta \left(-\frac{125}{192} h^{2} q^{2} + \frac{35}{42} h^{2} + \frac{31}{48} h q^{2} + \frac{3}{2} h + \frac{9}{8} q^{2}\right)
- \frac{3}{64} h^{2} q^{2} + \frac{1}{8} h^{2} + \frac{1}{16} h q^{2} + \frac{1}{2} h + \frac{1}{8} q^{2} . \nonumber \\
\end{eqnarray}
Similarly, the energy density can be calculated.
The results up to first order in $\beta$ and second order in $h,q$ are as follows:
\begin{eqnarray}
\frac{E}{N_{\rm site}} = \beta \left( -\frac{183}{64} h^2q^2 - \frac{189}{16} h q^2 - \frac{33}{8} h^2 + \frac{33}{8} q^2 -\frac{3}{2} \right) - \frac{9}{32}h^2q^2 -\frac{9}{8}h q^2 -\frac{3}{4}h^2 + \frac{3}{4}q^2 .
\nonumber \\
\end{eqnarray}

\end{document}